\documentclass[12pt]{article}
\def\a{\alpha}
\def\b{\beta}
\def\g{\gamma}
\def\vt{\vartheta}
\def\d{\delta}
\def\vta{\vartheta}

\def\cA{{\cal A}}
\def\cF{{\cal F}}
\def\cJ{{\cal J}}

\usepackage{amsmath}
\usepackage{amssymb,amsthm,graphicx,citesort}
\usepackage[ansinew]{inputenc}
\usepackage{stmaryrd}
\usepackage{multibox}
\usepackage{graphicx, epic, eepic, color}% Include figure files
\usepackage{dcolumn}% Align table columns on decimal point
\usepackage{bm}% bold math
\usepackage[ansinew]{inputenc}
\usepackage{times}

\begin{document}

\title{An assessment of Evans' unified field
  theory I}

\author{Friedrich W.~Hehl\footnote{Institute for Theoretical Physics,
    University at Cologne, 50923 K\"oln, Germany and Department of
    Physics and Astronomy, University of Missouri-Columbia, Columbia,
    MO 65211, USA}}

\date{2 July 2007, {\it file AssessI08.tex, fwh}}

\maketitle

\begin{abstract}
  Evans developed a classical unified field theory of gravitation and
  electromagnetism on the background of a spacetime obeying a
  Rie\-mann-Cartan geometry. This geometry can be characterized by an
  orthonormal coframe $\vt^\a$ and a (metric compatible) Lorentz
  connection $\Gamma^{\a\b}$. These two potentials yield the field
  strengths torsion $T^\a$ and curvature $R^{\a\b}$. Evans tried to
  infuse electromagnetic properties into this geometrical framework by
  putting the coframe $\vt^\a$ to be proportional to four extended
  electromagnetic potentials ${\cal A}^\a$; these are assumed to
  encompass the conventional Maxwellian potential $A$ in a suitable
  limit. The viable Einstein-Cartan(-Sciama-Kibble) theory of gravity
  was adopted by Evans to describe the gravitational sector of his
  theory. Including also the results of an accompanying paper by
  Obukhov and the author, we show that Evans' ansatz for
  electromagnetism is untenable beyond repair both from a geometrical
  as well as from a physical point of view. As a consequence, his
  unified theory is obsolete.
\end{abstract}

\begin{footnotesize}

\end{footnotesize}

\newpage
\section{Introduction}

One of the problems in evaluating Evans' unified field theory is that
its content is spread over hundreds of pages in articles and books of
Evans and his associates. There is no single paper in which the
fundamentals of Evans' theory are formulated in a concise and complete
way. Nevertheless, we can take Evans' papers
\cite{Evans2003,Evans2005a}, which subsume also work done earlier, as
a starting point. Now and then, when additional information is
required, we will use other publications of Evans and collaborators,
too \cite{workshop,EvansBookI,EvansBookII,EvansBookIII,EvansNo63}. We
will try to put the fundamental equations of Evans' theory in a way as
condensed as possible; in fact, we will come up with the nine
equations from (\ref{torsion*}) to (\ref{Sc2*}) that characterize
Evans' theory.  Incidentally, we came across Evans' unified field
theory in the context of a refereeing process. And in the present
paper, we will formulate our assessment in considerable detail.

We use, as Evans does in \cite{Evans2005a}, the calculus of exterior
differential forms. A translation for Evans' notation into ours is
given in Table \ref{translation} on the next page.

The evaluation of Evans' theory is made more demanding since his
articles contain many mathematical mistakes and inconsistencies, as
has been amply shown by Bruhn
\cite{Bruhn1,Bruhn2,Bruhn3,Bruhn4,Bruhn5,Bruhn6,BruhnBianchi,Bruhn7,Bruhn8}
and Rodrigues et al.\ \cite{Rodrigues1,Rodrigues2}. Let me just
illustrate this point with two new examples. I take Evans' ``Einstein
equation'' in \cite{Evans2005a}, App.4, Eq.(11), namely
$R_b^a=kT_b^a$. According to Evans' definition \cite{Evans2005a},
Eq.(16), the left hand side represents the curvature 2-form in a
Riemann-Cartan geometry (i.e., $R^{\a\b}=-R^{\b\a}$) and the right
hand side is proportional to the components of the canonical
energy-momentum tensor $T_b^a$.  Clearly, this equation is incorrect
since a 2-form $R_b^a=R_{\mu\nu b}^a\, dx^\mu\wedge dx^\nu/2$ with its
36 components cannot be equated to the 16 component of a second rank
tensor. If we generously interpreted $R_b^a$ as Ricci tensor, even
though Evans denotes the Ricci tensor always as $R_{\mu\nu}$, the
equation would be wrong, too, since on the left-hand-side of the
Einstein equation we have the Einstein and not the Ricci tensor. A
second example, we can find nearby: In \cite{Evans2005a}, App.4,
Eq.(10), Evans claims that the energy-momentum of his generalized
Einstein equation obeys $D\wedge T_b^a=0$. It is well-known, however,
that in a spacetime with torsion there can be no zero on the
right-hand-side, rather torsion and curvature dependent terms must
enter, see \cite{RMP}, Eq.(3.12).  Similar examples can be found
easily.

\renewcommand{\arraystretch}{1.4} % 2
\begin{table}[ht]\label{translation}
\begin{center}%\centering
\begin{tabular}{|c||c|c|}\hline
  Notion & Evans & here\\
  \hline coframe &$q^a=q^a_\mu dx^\mu$ & $\vt^\a =e_i{}^\a dx^i$\\ \hline
  connection & $\omega^a_b=\omega^a_{\mu b}dx^\mu$ & $\Gamma_\a{}^\b= 
  \Gamma_{i\a}{}^\b dx^i$\\
  \hline torsion & $T^a=\frac 12\, T^a_{\mu\nu}dx^\mu\wedge dx^\nu$ &
  $T^\a=\frac 12\, T_{ij}{}^\a dx^i\wedge dx^j$ \\ \hline curvature & $R^a_b=
  \frac 12\,R^a_{b\mu\nu}dx^\mu\wedge dx^\nu$ & $R_\a{}^\b=\frac
  12\,R_{ij\a}{}^\b dx^i\wedge dx^j$\\ \hline
  Ricci tensor/1-form& $R_{\mu\nu}$ & $ {\rm Ric}_\a=e_\b \rfloor
  R_\a{}^\b= {\rm Ric}_{\b\a}\vt^\b
  $ \\ \hline
  Evans' elmg.\ potential &$A^a$& ${\cal A}^\a$\\ \hline Evans' elmg.\
  constant& ${\cal A}^{(0)}$ &$a_0$\\ \hline Evans' elmg.\ field 
  strength & $F^a$ & ${\cal F}^\a$ \\  \hline Evans' hom.\ 
  current& $j^\nu$& $\cJ_{\rm hom}^\a$ \\ \hline Evans' inh.\
  current& $J^\nu$& $\cJ^\a_{\rm inh}$ \\ \hline can.\ energy-mom.\ density 
  & $T^a_b$ & ${\Sigma}_{\a}=\frak{T}_\a{}^\b\eta_\b$ \\ \hline spin
  ang.\ mom.\ density&?&$ \tau_{\a\b}=\frak{S}_{\a\b}{}^\g\eta_\g$\\ 
  \hline Hodge duality  &$\widetilde{\Psi}^a$&$^\star\Psi^\a$\\ \hline
\end{tabular}
\end{center}\medskip
\caption{Translation of Evans' notation into ours. Note that
 $\eta_\a={}^\star \vt_\a$. In Evans' work, $T^a_b$ is also sometimes
 used as symmetric energy-momentum tensor.}
%\bigskip
\end{table}

One may argue, as I will do in future, that a scientist educated as
chemist may have a great idea in physics even if the mathematical
details of his articles are not quite sound. Accordingly, I sometimes
followed not only that subclass of Evans' formulas that deemed correct
to me, but also his prose in oder to understand Evans' underlying
``philosophy''.

It is clear from \cite{Evans2005a} that the 4-dimensional spacetime in
which Evans' theory takes place obeys a Riemann-Cartan geometry
(RC-geometry) \cite{RMP} or, in the words of Evans, a
``Cartan-geometry". We decided to take what Evans calls an
antisymmetric part of the metric $\vt^{\a\b}:=\vt^\a\wedge\vt^\b$
(here $\vt^\a$ is the coframe of the RC-spacetime) not seriously as a
part of the metric, see Bruhn \cite{Bruhn6} for a detailed
investigation. The quantity $\vt^{\a\b}=-\vt^{\b\a}$ is an
antisymmetric tensor-valued 2-form with 36 independent components and
it is a respectable and useful quantity in RC-geometry, but it
certainly cannot be interpreted as part of a metric. Since Evans very
often claims that he uses a RC-geometry for the description of
spacetime, we take his word for it. Then, an additional geometric
structure, like an additional antisymmetric part of the metric is
ruled out.

In a RC-geometry, a linear connection $\Gamma$ and a metric $g$ with
Minkowskian signature $(+--\,-)$ are prescribed, furthermore metric
compatibility is required. This guarantees that {\it lengths and
  angles} are integrable in RC-geometry. Evans arrives at a
RC-geometry by means of what he calls the ``tetrad postulate'', see
\cite{Evans2005a}, Eqs.(32) and (33), and Rodrigues et al.\
\cite{Rodrigues1}.

In Sec.2 we will display the geometric properties of a RC-geometry in
the 4-dimensional spacetime in quite some detail. In particular, we
define torsion and curvature and decompose torsion into components
belonging to different irreducible representation spaces of the
Lorentz group. We introduce the contortion and the Ricci 1-forms and
the curvature scalar. Moreover, the two Bianchi identities are
displayed and two irreducible pieces projected out leading to the
Cartan and Einstein 3-forms. The Ricci identity will be mentioned
shortly.

In Sec.3 we will take Evans' ansatz relating the coframe $\vt^\alpha$
to a generalized electromagnetic potential ${\cal A}^\alpha$ according
to ${\cal A}^\alpha=a_0\,\vt^\alpha$, where $a_0$ is a scalar factor
of the dimension {\it magnetic flux /length\/} $\stackrel{\rm SI}{=}
W\!b/m=Vs/m$, see Evans and Eckardt \cite{EvansNo63}, p.2.  We will
point out that in this way one finds four Lorentz-vector valued
1-forms or, in other words, extended electromagnetic
$SO(1,3)$-covariant potentials ${\cal A}^\alpha$ with 16 components,
in contrast to what Evans finds, namely $O(3)$-covariant potentials.
Then the extended electromagnetic field strength $\cF^\a$ is defined
and the generalized Maxwell equations displayed and discussed.

In Sec.4 we show that Evans just adopted the viable Einstein-Cartan
theory (EC-theory) of {\em gravity} for his purpose literally. His
generalized Einsteinian field equation is the same as the first field
equation of EC-theory. As a consequence of the angular momentum law,
Evans also used the second field equation of EC-theory, even though he
used it only in words in identifying torsion with the spin of matter.
Then we display the energy-momentum and angular momentum laws. It is
pointed out that the so-called Evans wave equation for the coframe
$\vt^\a$ is a {\it redundant} structure since the dynamics of $\vt^\a$
is already controlled by the generalized field Einstein equation of
EC-theory together with the corresponding Cartan field equation of
gravity.

In Sec.5 we collect the fundamental equations of Evans' theory in the
nine equations from (\ref{torsion*}) to (\ref{Sc2*}). We will explain
what exactly we call Evans' unified field theory. In an accompanying
paper by Obukhov and the author \cite{AssII}, we propose a new
variational principle for Evans' theory and derive the corresponding
field equations of Evans' theory. It turns out that for all physical
cases we can derive the vanishing of torsion and thus the collapse of
Evans' theory to Einstein's ordinary field equation.  We discuss our
findings, including the results of \cite{AssII}, and summarize our
objections against Evans' unified theory.

A few historical remarks may be in order. Cartan himself noticed in a
letter to Einstein, see \cite{Debever}, page 7, that one irreducible
piece of the torsion $T$ ``has precisely all the mathematical
characteristics of the electromagnetic potential''; it is apparently
the vector piece $T_{\rm vec}\sim e_\a\rfloor T^\a$ that he determined
earlier, between 1923 and 1925, in \cite{Cartan}.  Thus he discussed
$T_{\rm vec}\sim A$, where $A$ is the potential of Maxwellian
electrodynamics. Note that this assumption is totally different from
Evans' ansatz $\vt^\a\sim \cA^\a$.  Moreover, Cartan did {\it not}
develop a corresponding electromagnetic theory.  In fact, in the same
papers \cite{Cartan}, he linked, within a consistent theoretical
framework, torsion to the spin of matter. He laid the groundwork to
what we call nowadays the EC-theory of gravity
\cite{Blagojevic,RMP,Trautman}.  This excludes the mentioned
identification of a piece of the torsion with the electromagnetic
potential.

Later Eyraud \cite{Eyraud} and Infeld \cite{Infeld} and, more
recently, Horie \cite{Horie} tried to link torsion to the
electromagnetic field. But these attempts did lead to nowhere. For
more details, one may consult Tonnelat \cite{Tonnelat} and Goenner
\cite{Goenner}.  \medskip

{\it Note added after the revision of the paper:} In the meantime,
Evans on 28 March 2007 put a revised rebuttal on the net in which he
tries to answer my objections.\footnote{See
  http://www.aias.us/documents/rebuttals/ahehlrebuttal.pdf .} I
studied this rebuttal carefully and improved my presentation here and
there in order to make my arguments more comprehensible. I also
changed some ambiguous statements. However, I found no counterargument
of Evans really convincing. Evans also complains that I referred
extensively to the work of Bruhn, but that I did not mention his
answers. This has a simple reason: Bruhn found many mathematical
mistakes in Evans' work, but instead of improving his computations
correspondingly, Evans just bluntly rejects all of Bruhn's correct
results as irrelevant in a more than offensive language. I feel no
need to quote such outpourings of anger.

In Sec.5.1, I reduced the foundations of Evans' theory to the {\em
  nine equations} (\ref{torsion*}) to (\ref{Sc2*}) and in Sec.5.2, I
listed {\em five cornerstones} of Evans' theory that, in my opinion,
lie at the foundations of Evans' theory. Nowhere in Evans' rebuttal he
says anything about what I consider to be the foundation of his
theory.  Is he not sure about these foundations? I can only understand
Evans' silence in this respect in two ways: Either he doesn't know the
answer or he does not want to commit himself to a definite structure
of his theory and prefers to leave his theory in mystic darkness. 

Let me just give an example. An associate of Evans, with whom I
discussed a lot, firmly believes that cornerstone 4 (the generalized
Einstein equation) does {\em not} belong to Evans' unified field
theory.  However, in Evans' paper \cite{EvansBookI} this equation is
postulated and it takes a central part in Evans'
flowcharts\footnote{See
  http://www.aias.us/index.php?goto=showPageByTitle\&pageTitle=Equations\underline{\,\null\,}flowcharts
  .} of 1 July 2007 (see the so-called ``Evans field equation'').
Accordingly, the situation has not changed after Evans' rebuttal:
Somebody who wants to learn about the structure of Evans' theory has
to turn to my paper, be it a follower of Evans or just somebody who
wants to know it.  From Evans' numerous articles this information
cannot be extracted so easily.\medskip

I invite Evans to scrutinize\vspace{-8pt}
\begin{enumerate} 
\item  our nine equations (\ref{torsion*}) to (\ref{Sc2*}) and\vspace{-8pt}
\item our five cornerstones in Sec.5.2.
\end{enumerate}
If he finds our description of the foundations of his theory
incorrect, he should state this explicitly and should specify his own
cornerstones on one or two sheets of paper and should compare them
with what I found. Just to cite some 20 of his papers is not helpful.
Evans' followers as well as critical scientists are interested in the
display of these foundations. Only then one can judge a theory.

\section{Geometry: Riemann-Cartan geometry of spacetime}

\subsection{Defining RC-geometry}

We assume a 4-dimensional differential manifold. At each point, the
basis of the tangent space are the four linearly independent {\it
  vectors} $\mathbf{e}_\a=e^i{}_\a\partial_i\,$, here
$\a,\b,...=0,1,2,3$, the (anholonomic) tetrad indices, number the
vectors and $i,j,k,...=0,1,2,3$, the (holonomic) coordinate indices,
denote the components of the respective vectors. The basis of the
cotangent space is span by the four linearly independent {\it
  covectors} or 1-forms $\mbox{\boldmath$\vartheta$\unboldmath}^\b
=e_j{}^\b\,dx^j\,$. The bases of vectors and covectors are dual to
each other.  Consequently, we have $e^i{}_\a\, e_i{}^\b=\d^\b_\a\,$
and $e^i{}_\b\,e_j{}^\b=\d_j^i$. We call collectively the $e_\a$'s and
the $\vt^\b$'s also {\it tetrads}. We follow the
conventions\footnote{\;We build from the coframe $\vt^\a$ by exterior
  multiplication and by applying the Hodge star the following
  expressions: $ \vt^{\a\b}:=\vt^\a\wedge\vt^\b\,,\;
  \vt^{\a\b\g}:=\vt^{\a\b}\wedge \vt^\g\,,\;\mbox{etc.;}\quad
  \eta:={}^\star 1\>\;\mbox{(volume 4-form)}\,,\quad
  \eta^\a:={}^\star\vt^\a\,,\;
  \eta^{\a\b}:={}^\star\vt^{\a\b}\,,\;\mbox{etc.}$, see
  \cite{PRs,Birkbook}.  We denote antisymmetrization by brackets $[ij]
  = (ij -ji)/2$ and symmetrization by parentheses: $(ij)=(ij+ji)/2$.
  Analogously for more indices, as, e.g., for $[ijk]$, where we have
  $[ijk]=(ijk-jik+jki-+\cdots)/3!$, see Schouten
  \cite{Schouten,Schouten*}.  } specified in \cite{PRs}.

On our manifold we impose a {\it connection} 1-form
$\Gamma_\a{}^\b=\Gamma_{i\a}{}^\b dx^i$ that allows us to define the
parallel transport of quantities. In particular, for the frame $e_\a$,
we have $De_a=\Gamma_\a{}^\b e_\b$. For an arbitrary tensor-valued
form, the {\it covariant exterior derivative} operator is
$D:=d+\Gamma_\a{}^\b\,f^\a{}_\b\,$, where $f^\a{}_\b$ represents the
behavior of the quantity under linear transformations of the frames.
Additionally, we impose a symmetric {\it metric}
$g=g_{ij}\,dx^i\otimes dx^j$, with $g_{ij}=g_{ji}$. Referred to a
tetrad, we have $g_{\a\b}=e^i{}_\a e^j{}_\b\,g_{ij}$.

Metric and connection are postulated to be compatible, that is, the
nonmetricity 1-form $Q_{\a\b}:=-Dg_{\a\b}$ is postulated to vanish:
$Q_{\a\b}=0$. This guarantees that lengths and angles are constant
under parallel transport. In accordance with this fact, it is
convenient to choose the tetrads to be {\it orthonormal} once and for
all. Then, $g_{\a\b}=\mbox{diag}(+1,-1,-1,-1)=:o_{\a\b}$, where
$o_{\a\b}$ is the Minkowski metric. If we raise the $\a$-index of the
connection, then $\Gamma^{\a\b}=-\Gamma^{\b\a}$. This is known as the
Lorentz (or spin) connection. The one--form $\Gamma$ takes values in
the Lie algebra of $SO(1,3)$. Hence, the variables $\vt^\a$ and
$\Gamma^{\a\b}$, i.e., coframe and Lorentz connection, specify the
geometry completely.

In the subsequent section, we need to discuss the transformation
properties of the coframe $\vt^\a=e_i{}^\a dx^i$. Under a coordinate
transformation, it behaves like a 1-form, in components,
$e_{i'}{}^\a=\frac{\partial x^j}{\partial x^{i'}}\,e_j{}^\a$.  Under
local SO(1,3) Lorentz rotations $\Lambda_\b{}^{\a'}$, it transforms as
a Lorentz vector $\vt^{\a'}=\Lambda_\b{}^{\a'}\,\vt^\b$. Similarly,
the frame $e_\b=e^j{}_\b\partial_j$ transforms as
$e^{j'}{}_\b=\frac{\partial x^{j'}}{\partial x^k}\,e^k{}_\b$ under
coordinate and as $e_{\b'}=\Lambda_{\b'}{}^\g e_\g$ under Lorentz
transformations, with $\Lambda_{\b'}{}^\g\Lambda_\g{}^{\a'}=
\delta^{\a'}_{\b'}$. The spatial rotation group $O(3)$ is a subgroup
of $SO(1,3)$. But the Lorentz group includes also the boosts. In other
words, whereas a spatial $O(3)$-rotation of $\vt^\a$ is an allowed
procedure, the theory is only locally Lorentz covariant --- and thus
takes place in a RC-geometry --- if the coframe $\vt^\a$ transforms
under the complete $SO(1,3)$. 

The geometry defined so far is called RC-geometry. It is also clear
from the statements of {\it Evans} that he uses {\it exactly the same
  geometry}. Thus, we have a secure platform for our evaluation. 

Incidentally, in four dimensions, RC-geometry was first applied in the
viable Einstein-Cartan(-Sciama-Kibble) theory of gravity, for short
EC-theory
\cite{Sciama1,Kibble,Sciama2,Trautman73,RMP,Kinematics,Trautman}.

\subsection{Torsion and curvature}

From our variables $\vt^\a$ and $\Gamma_a{}^\b$, we can extract two
Lorentz tensors, the {\it torsion} and the {\it curvature} 2-forms,
respectively:
\begin{eqnarray}\label{tor}
 T^\a&:=&D\vt^\a=d\vt^\a+\Gamma_\b{}^\a\wedge \vt^\b\,,\\ \label{curv}
  R_\a{}^\b&:=&d\Gamma_\a{}^\b-\Gamma_\a{}^\g\wedge\Gamma_\g{}^\b\,.
\end{eqnarray}
In RC-geometry, we have, because of the metric compatibility,
$\Gamma^{\a\b}=-\Gamma^{\b\a}$, and thus, $R^{\a\b}=-R^{\b\a}$. In
four dimensions, torsion and curvature have 24 and 36 independent
components, respectively.

If one applies a Cartan displacement (rolling without gliding) around
an infinitesimal loop in the manifold, then {\it torsion} is related
to the {\it translational} misfit and the curvature to the rotational
misfit; a discussion can be found in Cartan's lectures
\cite{CartanGoldberg}, see also Sharpe \cite{Sharpe}, our book
\cite{Birkbook}, Sec.C.1.6, and the recent article of Wise
\cite{Wise}.  Alternatively, one may build up an infinitesimal
parallelogram, then the {\it translational closure failure is
  proportional to the torsion}, see Bishop and Crittenden
\cite{Bishop}, p.97. This is important: Geometrically, from the point
of view of RC-geometry, torsion has nothing to do with spin, but
rather with translations. It is for this reason that torsion can be
understood as the field strength of a translational gauge theory, see
Pilch \cite{Pilch} and Gronwald \cite{Gronwald}.  Consequently, when
Evans treats torsion and spin (Spin of matter? Spin of gravity? Spin
of electromagnetism?)  synonymously, as he does in all of his articles
on his theory,\footnote{\;``There are two fundamental differential
  forms...that together describe any spacetime, the torsion or spin
  form and Riemann or curvature form.''  See Evans \cite{Evans2005a},
  p.434.  Just by the choice of Evans' language, torsion is always
  identified with spin. We are not told what sort of spin we have to
  think of. In Sec.4 we will see that it has to be the total spin of
  all matter and the electromagnetic field, with exception of
  gravity.}  then this can only be understood as an additional {\it
  dynamical assumption} that is independent from the RC-geometry of
the underlying spacetime.\footnote{In his rebuttal, see footnote 1,
  Evans argues in the following way: Let us write (\ref{tor}) in terms
  of components, namely
  $T_{ij}^\a=2\left(\partial_{[i}e_{j]}{}^{\a}+\Gamma_{[i|\b}^\a
    e_{j]}{}^{\b} \right)$. This equation ``clearly has the
  anti-symmetry needed for angular momentum and torque. These
  quantities involve spin and this is meticulously defined in ECE
  theory ... in many places.'' Angular momentum and torque are
  quantities defined in {\it mechanics} and in {\em field theory,}
  respectively, and cannot be taken from geometry generally or from
  RC-geometry specifically. That the antisymmetry of a certain
  geometric quantity, here torsion, reminds of angular momentum and
  torque is one thing, to really interrelate spin angular momentum
  with torsion is an additional dynamical assumption that does not
  follow from RC-geometry alone. In Sec.5.2, we call this assumption
  of Evans ``cornerstone five'' of his theory.}  We will see further
down in detail that this is, indeed, the case.

The torsion 2-form $T^\a$ can be contracted
by $e_\a\rfloor$ to a covector $e_\a\rfloor T^\a$ and multiplied by
$\vt_\a$ to yield a 3-form $\vt_\a\wedge T^\a$ or, using the Hodge
star, to a covector with twist $^\star\left(\vt_\a\wedge T^\a\right)$.
These expressions correspond to the vector and the axial vector pieces
of the torsion.  More formally, we can decompose the torsion tensor
irreducibly under the local Lorentz into three pieces:
\begin{equation}\label{tordecomp}
T^\a={}^{(1)}T^\a+{}^{(2)}T^\a+{}^{(3)}T^\a\,.
\end{equation}
The second and the third pieces correspond to the mentioned {\it
  vector} and {\it axial vector} pieces, respectively,
\begin{eqnarray}\label{vecT}
  ^{(2)}T^\a &:=&  \frac 13\,\vt^\a\wedge (e_\b\rfloor T^\b)\,,\\
  ^{(3)}T^\a &:=& \frac 13\,e^\a\rfloor(\vt_\b\wedge T^\b)\,,
  \label{axiT}         
\end{eqnarray}
whereas the first piece can be computed by using (\ref{tordecomp}).

For a comparison with Riemannian geoemtry, it is often convenient to
decompose the connection 1-form into a Riemannian part, denoted by a
tilde, and a tensorial post-Riemannian part according to
\begin{equation}\label{postR}
\Gamma_\a{}^\b=\widetilde{\Gamma}_\a{}^\b -K_\a{}^\b\,. 
\end{equation}
In RC-geometry, the $K_\a{}^\b$ can be derived by evaluating
$Dg_{\a\b}=0$. We find the {\it contortion} 1-form as \cite{PRs}
\begin{equation}\label{contortion}
  K_{\alpha\beta}:= 2e_{[\alpha}\rfloor T_{\beta ]}
  -\frac 12\, e_{\alpha}\rfloor e_{\beta}\rfloor
  (T_{\gamma}\wedge\vartheta^{\gamma})=-K_{\beta\alpha}\,.
\end{equation}
Resolved with respect to the torsion, we have $T^{\alpha} =
K^{\alpha}{}_{\beta}\wedge \vartheta^{\beta}$.

The curvature 2-form yields, by contraction, the Ricci 1-form
\begin{equation}\label{ricci1} 
{\rm Ric}_\alpha := e_\beta\rfloor R_\alpha{}^\beta =
  {\rm Ric}_{\beta\alpha}\,\vartheta^\beta=R_{\g\b\a}{}^\g\vt^\b.
\end{equation}
In a RC-geometry, the components of the Ricci 1-form are asymmetric in
general: ${\rm Ric}_{\a\b}\ne {\rm Ric}_{\b\a}$. By transvection with
the metric, we find the curvature scalar 
\begin{equation}\label{scalar}
  R:=g^{\a\b}\,{\rm Ric}_{\a\b}=g^{\a\b}e_\a\rfloor {\rm Ric}_\b=
e_\a\rfloor e_\b\rfloor R^{\a\b}
  =-e_\a\rfloor\left[e_\b\rfloor\,^\star\!\left({}^\star\! R^{\a\b}\right)
  \right] \,. 
\end{equation}
After some algebra, see \cite{Birkbook}, p.338, we find for the
curvature scalar, with $\eta_{\a\b}=\,^\star\left(\vt_\a\wedge\vt_\b
\right)$, the following:
\begin{equation}\label{scalar'}
R=e_\a\rfloor e_\b\rfloor R^{\a\b}=\,^\star\!
\left(\eta_{\a\b}\wedge R^{\a\b} \right)\,.
\end{equation}
The expression under the star can be taken as a Lagrangian 4-form of
the gravitational field.

The curvature 2-form can be decomposed into 6 different pieces, see
\cite{PRs}. Among them, we find the symmetric tracefree Ricci tensor,
the curvature scalar, and the antisymmetric piece of the Ricci tensor.

\subsection{Bianchi identities}

If we differentiate (\ref{tor}) and (\ref{curv}), we find the two
Bianchi identities for torsion and curvature, respectively:
\begin{eqnarray}\label{Bianchi1}
  DT^\a&=&R_\b{}^\a\wedge\vt^\b\,,\\ DR_\a{}^\b&=&0\,.\label{Bianchi2}
\end{eqnarray}
Incidentally, Evans agrees that he and we use the same
RC-geometry.\footnote{M.W. Evans states in his internet blog
  http://www.atomicprecision.com/blog/2006/12/08/
  endorsement-of-ece-by-the-profession/ the following: ``The two
  Cartan structure equations, two Bianchi identities and tetrad
  postulate used by Carroll, Hehl and myself are the same.''  See,
  however, a note of Bruhn \cite{BruhnBianchi} on some mistake in the
  corresponding considerations of Evans.}

Both Bianchi identities can be decomposed irreducibly under the local
Lorentz group into 3 and 4 irreducible pieces, respectively; for
details, see \cite{HMcCrea,PRs,McMapping}. We remind ourselves of the
1-form $\eta_{\a\b\g}=\,^\star\!\left(\vta_\a\wedge\vta_\b
  \wedge\vta_\g\right)$, where the star denotes the Hodge operator.
Then, by exterior multiplication of the Bianchi identities with this
1-form, we can extract from (\ref{Bianchi1}) an irreducible piece with
6 independent components,
\begin{equation}\label{Bianchi1irr}
  DT^\g\wedge\eta_{\g\a\b}=
R_\d{}^\g\wedge\vta^\d\wedge \eta_{\g\a\b}\,,
\end{equation}
and from (\ref{Bianchi2}) one with 4 independent components,
\begin{equation}\label{Bianchi2irr}
DR^{\b\g}\wedge\eta_{\b\g\a}= 0\,.
\end{equation}
We define the {\it Cartan\/} and the {\it Einstein\/} 3-forms,
\begin{eqnarray}\label{Car}
C_{\a\b}&:=&\frac 12\,\eta_{\a\b\g}\wedge T^\g\,,\\ \label{Ein}
G_\a&:=&\frac 12\,\eta_{\a\b\g}\wedge R^{\b\g}\,,
\end{eqnarray}
respectively. Now we shift in (\ref{Bianchi1irr},\ref{Bianchi2irr}) by
partial integration the 1-form $\eta_{\a\b\g}$ under D. After some
algebra, we find,
\begin{eqnarray}\label{DC}
  DC_{\a\b}&=& -
\eta_{[\a}\wedge {\rm Ric}_{\b]}\,,\\ \label{DG}
DG_\a&=&\frac 12\,\eta_{\a\b\g\d}\,R^{\b\g}\wedge T^\d\,.
\end{eqnarray}
Thus, the Cartan 3-form $C_{\a\b}$ and the Einstein 3-form $G_\a$ are
important quantities since they appear in the two contracted Bianchi
identities (\ref{DC}) and (\ref{DG}) under the differentiation symbol.

Using (\ref{Ein}) and the formula $\vt_{[\a}\wedge G_{\b]}=
\eta_{[\a}\wedge {\rm Ric}_{\b]}$, Eq.(\ref{DC}) can be rewritten in
terms of $G_\a$. Thus, finally we have for the two contracted Bianchi
identities, see also \cite{Obukhov2006a,Obukhov2006b},
\begin{eqnarray}\label{DCar}
DC_{\a\b}+\vt_{[\a}\wedge G_{\b]}&=&0\,,\\ \label{DEin}
DG_\a&=&\frac 12\,\eta_{\a\b\g\d}\,R^{\b\g}\wedge T^\d\,.
\end{eqnarray}
We will come back to these $6+4$ independent equations below. Note
that (\ref{DEin}) is only valid in four dimensions. In {\it three\/}
dimensions --- then $G_\a$ is a 2-form --- the term on the
right-hand-side of (\ref{DEin}) vanishes,\footnote{\;\'E.~Cartan
  \cite{Cartan1922,cartan1922} worked very intuitively. One of his
  goals in analyzing Einstein's theory was to get a geometrical
  understanding of the Einstein {\it tensor}, that is, to get hold of
  the Einstein tensor without using analytical calculations. He
  achieved that for three dimensions, where $DG_\a=0$. Obviously
  Cartan's intuition worked with three dimensions. There is evidence
  for this, namely, he constructed a special 3-dimensional model of a
  RC-space \cite{Cartan1922,cartan1922}, the Cartan spiral staircase;
  for a discussion, see Garcia et al.\ \cite{Garcia}, Sec.V.
  Apparently over-stretching his intuition, Cartan also assumed
  $DG_\a=0$ for four dimensions and run into difficulties with his
  gravitational theory.  We go into such details here, since Evans
  \cite{EvansBookI}, p.\ 464, commits the same mistake as Cartan did
  and assumes $DG_\a=0$ for four dimensions, whereas, in fact,
  (\ref{DEin}) is correct. It can be taken from the contracted second
  Bianchi identity (\ref{DEin}) that postulating $DG_\a=0$ in four
  dimensions yields the constraint $\eta_{\a\b\g\d}\,R^{\b\g}\wedge
  T^\d=0$, a geometrical ad hoc assumption that has no motivation and
  restricts, on a purely kinematical level, the free choice of the
  torsion and of the curvature of the RC-geometry appreciably. The
  assumption $DG_\a=0$ in four dimensions is {\em not} part of
  RC-geometry. There rather (\ref{DEin}) is valid. The comparison
  between the four-dimensional EC-theory and a three-dimensional
  continuum theory of lattice defects has been reviewed by Ruggiero \&
  Tartaglia \cite{Ruggiero}.}  see \cite{HMcCrea}, that is, $DG_\a=0$
(for $\a=1,2,3$).

\subsection{Ricci identity}
If we take the exterior covariant derivative of the vector-valued
$p$-form $\Psi^\a$, we find, because of $dd=0$, the Ricci identity
\begin{equation}\label{RicciId}
DD\Psi^\a = R_\b{}^\a\wedge\Psi^\b\,.
\end{equation}
This is particularly true for the coframe,
\begin{equation}\label{RicciId'}
DD\vt^\a = R_\b{}^\a\wedge\vt^\b\,.
\end{equation}
Due to $T^\a=D\vt^\a$, this corresponds to the first Bianchi identity
(\ref{Bianchi1}).
%Evans is calling the Ricci identity (\ref{RicciId'}) simply ``Evans' 
%lemma'', see \cite{EvansBookI}, p.464. However, he made a
%calculational mistake and found for the right-hand-side of
%(\ref{RicciId'}) effectively the expression $R\,^\star\vt^\a$, instead
%of the correct $ R_\b{}^\a\wedge\vt^\b$.

\section{Electromagnetism: Evans' ansatz for extended 
electromagnetism}

\subsection{Evans' ansatz}\label{ELEKTRO}

Up to now, everything is quite conventional. The RC-geometry, which
Evans is using, has been introduced earlier in gauge theories of
gravity and is well understood, see \cite{Erice}. However, in the
electromagnetic sector, Evans turns to an ad hoc ansatz.  He assumes
the existence of an extended electromagnetic potential $\cA^\a$ that
is proportional to the coframe $\vt^\a$,
\begin{equation}\label{ansatz}
{\cal A}^\a=a_0\,\vt^\a\,\qquad\mbox{or}\qquad \cA_i{}^\a=a_0\,e_i{}^a\,,
\end{equation}
see Evans \cite{Evans2005a}, Eq.(12). Here $a_0$ denotes a scalar
constant of dimension $[a_0]=[\cA^\a]/[\vt^a]=$ {\it magnetic
  flux\,/length;} it has supposedly to be fixed by experiment.

Due to the omnipresence of the coframe $\vt^\a$ (apart from singular
points, see Jadczyk \cite{Ark}, but also Tresguerres and Mielke
\cite{RomualdoEgg}), an extended electromagnetic potential $\cA^\a$ is
created by (\ref{ansatz}) everywhere. Thus, one may call such an
ansatz pan-electromagnetic. The constant $a_0$ must be thought of as a
universal constant. Otherwise a geometric theory, which supposedly
describes a universal interaction, looses its raison d'{\^e}tre. The
dimension of $a_0$ doesn't point to its universality. Remember that
universal constants usually have the dimensions of $
q^{n_1}\,\frak{h}^{n_2}$, where $q$ denotes the dimension of a charge
and $\frak{h}$ that of an action, see Post \cite{Post} and
\cite{Okun}. Constants built according to this rule, are 4-dimensional
scalars, since $q$ and $\frak{h}$ carry exactly this property.
Observationally it turns out that $n_1$ and $n_2$ are integers.
Examples for such dimensionful 4-scalars are
\begin{equation}\label{examplesSC}
  q\rightarrow\mbox{electric charge}\,,\>\; \frac{\frak{h}}{q}\rightarrow
  \mbox{magnetic flux}\,, \>\;
  \frac{\frak{h}}{q^2}\rightarrow\mbox{electric resistance}\dots
\end{equation}
Thus, $n_1,n_2=0,\pm1,\pm2,\dots$. Accordingly, the impedance of free
space $\Omega_0=\sqrt{\mu_0/\varepsilon_0}$, for example, is a
4-dimensional scalar and a universal constant, whereas $\varepsilon_0$
and $\mu_0$ for themselves are no 4-scalars.  And for $a_0$, we have
$[a_0]=\frak{h}/(q\times \mbox{\it length})$. This doesn't smell particularly
universal. The constant $a_0$ is not expected to qualify as a
4-scalar, since it defies the scheme (\ref{examplesSC}).

Evans has the following to say\footnote{\;We denote Evans' constant
  $A^{(0)}$ by $a_0$, see our Table 1.} (\cite{Evans2005a}, p.435):
``Here $A^{(0)}$ denotes a $\hat{C}$ negative scalar originating in
the magnetic fluxon $\hbar/e$, a primordial and universal constant of
physics.''  From \cite{Evanspaper55}, p.2 we learn that we have ``...a
scalar factor $A^{(0)}$, essentially a primordial voltage.'' In fact,
the dimension of $a_0$ is neither that of a magnetic flux nor that of
a voltage, but rather {\it magnetic flux/length}. The argument that a
universal constant should be a four dimensional scalar is very
suggestive to us, but it cannot be considered as conclusive.

For convenience we can parametrize $a_0$ with the help of the magnetic
flux quantum $h/(2e)$. Here $h$ is the Planck constant and $e$
the elementary charge. Then,
\begin{equation}\label{Elength}
a_0=\frac{h}{2e\ell_{\rm E}}\,.
\end{equation}
Thus, the length $\ell_{\rm E}$, the $E$ stands for Evans, is the new
unknown constant, which, incidentally, is nowhere defined in Evans'
publications. According to Evans, $a_0$ should be negative. Then the
same is true for $\ell_{\rm E}$.

The extended electromagnetic potential ${\cal A}^\a$ is represented by
four 1-forms,
\begin{eqnarray}\label{1-forms}
  \cA^0=\cA_i{}^0\,dx^i\,,\>\;\cA^1=\cA_i{}^1\,dx^i\,,\>\;
  \cA^2=\cA_i{}^2\,dx^i\,,\>\;
  \cA^3=\cA_i{}^3\,dx^i\,.
\end{eqnarray}
Thus, it has 16 independent components, quite a generalization as
compared to the Maxwellian potential $A=A_i\,dx^i$ with only 4
independent components. Evans doesn't give a Lorentz covariant
prescription of how to extract from $\cA^\a$ the Max\-wellian potential
$A$.  According to (\ref{ansatz}), ${\cal A}^\a$ transforms under a
local {\it Lorentz transformation} $\Lambda_\b{}^{\a'}$ like the
coframe:
\begin{equation}\label{transform}
\cA^{\a'}=\Lambda_\b{}^{\a'}\cA^\b\,.
\end{equation}
Suppose we try to identify the Maxwellian $A$ with $\cA^0$. Then,
under a local Lorentz rotation of the frame, this identification is
mixed up:
\begin{equation}\label{identify}
  \cA^{0'}=\Lambda_\b{}^{0'}\cA^\b=
  \Lambda_0{}^{0'}\cA^0+\Lambda_1{}^{0'}\cA^1+
  \Lambda_2{}^{0'}\cA^2 +\Lambda_3{}^{0'}\cA^3\,.
\end{equation}
In the new frame, indicated by a prime, $\cA^{0'}$ cannot be
identified with $A$ since it contains three non-Maxwellian admixtures.
However, for the physical description the new frame is {\it
  equivalent} to the old one. In other words, {\it the identification
  of $\cA^0$ as Maxwellian potential is not Lorentz covariant and has
  to be abandoned.} This is an inevitable consequence of the fact that
$\cA^\a$ transforms as a vector under the Lorentz group $SO(3,1)$, as
it does, according to Evans' ansatz (\ref{ansatz}). Similar
considerations apply to $\cA^1$, $\cA^2$, and $\cA^3$.

One could try to eliminate the $\a$-index in $\cA^\a$ by some
contraction procedures, such as $\vt_\a\wedge\cA^\a$ or
$e_\a\rfloor\cA^\a$; however, the former yields a 2-form, the latter a
0-form. Also the Hodge star doesn't help, since $^\star\vt_\a\wedge
\cA^\a$, e.g., represents a 4-form. Since Maxwell's theory in a
RC-spacetime is locally Lorentz covariant, the extraction of Maxwell's
potential 1-form $A$ from $\cA^\a$ doesn't seem to be possible.

Evans also considers 3-dimensional spatial rotations
$\rho_\b{}^{\a'}$. The corresponding rotation group $O(3)$, is a
subgroup of the Lorentz group $SO(1,3)$. Hence we can study the
behavior of $\cA^\a$ under these rotations:
\begin{equation}\label{rotation}
\cA^{\a'}=\rho_\b{}^{\a'}\cA^\b\,.
\end{equation}
This equation is contained in (\ref{transform}), which, additionally,
encompasses {\it boosts} in three linearly independent directions.
Clearly, the $O(3)$ is not the covariance group of $\cA^\a$. It is
just a subgroup of the $SO(1,3)$. An $O(3)$ covariant electromagnetic
potential cannot be derived from the ansatz (\ref{ansatz}) in a
Lorentz covariant way --- in contrast to what Evans claims
\cite{Evans2005a}.

Thus, instead of the desired $O(3)$-covariant extended electromagnetic
framework, Evans in fact, due to his ansatz (\ref{ansatz}),
constructed willy nilly a $SO(1,3)$-covariant framework. Still, he
insists that the $O(3)$-substructure has a meaning of its own;
however, certainly not in a Lorentz covariant sense.

If we differentiate Evans' ansatz, we find for the extended
electromagnetic field strength
\begin{eqnarray}\label{fieldF}
   {\cal F}^\a:=D{\cal A}^\a=d{\cal A}^\a
  +\Gamma_\b{}^\a\wedge {\cal A}^\b
\end{eqnarray}
the relation
\begin{equation}\label{ansatzF}
 {\cal F}^\a=a_0\,T^\a\,.
\end{equation}
Now we have $6\times 4$ components of the extended electromagnetic
field strength.  As we pointed out in Sec.2 and as it is known form
the literature, see, in particular Cartan \cite{CartanGoldberg} and
Bishop \& Crittenden \cite{Bishop}, the torsion $T^\a$ is a quantity
related to translations and, accordingly, to energy-momentum.  On the
left-hand-side, we have an extended electromagnetic quantity that is
eventually related to hypothetical extended electric currents.  Also
$\cF^\a$, like its potential $\cA^\a$, transforms as a vector under
Lorentz transformations:
\begin{eqnarray}\label{LorField}
\cF^{\a'}=\Lambda_\b{}^{\a'}\cF^\b\,.
\end{eqnarray}

Before we turn to the extended electromagnetic field equations of
Evans, let us first remind ourselves of the fundamental structure of
Maxwell's theory. We will follow here the premetric approach, see
\cite{Birkbook}, that separates the Maxwell equations from the
constitutive relation. We define first the four-dimensional electric
current 3-form $J=\rho-j\wedge dt$, with $\rho$ as charge density and
$j$ as current density. We postulate that charge is conserved for an
{\it arbitrary} four-dimensional domain of spacetime. This implies, in
particular, that $dJ=0$ and $J=dH$, with the electromagnetic excitation
2-form $H={\cal D}-{\cal H}\wedge dt$. By means of the Lorentz force
density $f_\a=\left(e_\a\rfloor F \right)\wedge J$, see
\cite{Birkbook}, one can define the electromagnetic field strength
2-form $F=B+E\wedge dt$. Postulating conservation of the magnetic flux
yields $dF=0$.  Accordingly, we found the inhomogeneous and the
homogeneous Maxwell euqations, respectively:
\begin{eqnarray}\label{Maxwell1}
 dH=J\,,\qquad dF=0\,.
\end{eqnarray}
These equations are generally covariant and are valid in this form in
special and in general relativity and in Riemann-Cartan spacetimes
likewise. In particular, they are free of the metric and of the
connection of spacetime. Both equations correspond to separate
physical facts, namely to charge and to flux conservation,
respectively, and are thus {\it independent} from each other.

In order to complete the theory, we have to specify, in addition to
the Maxwell equations, a {\it constitutive law\/}. In {\it vacuum},
that is, in free space without space charges, the field strength and
the excitation are related by
\begin{eqnarray}\label{constitutive}
H=\frac{1}{\Omega_0}\,^\star\!F\,,
\end{eqnarray}
where $\Omega_0=\sqrt{\mu_0/\varepsilon_0}$ is the impedance of free
space. The metric of spacetime is contained in the Hodge star. Now the
Maxwell equations for vacuum can be put into the form\footnote{\;For
  no obvious reason, Evans writes $d\,^\star\!F=\mu_0\,J$ instead.
  Apparently the $\varepsilon_0$ got lost.}
\begin{eqnarray}\label{Maxwell2}
dF=0\,,\qquad d\,^\star \! F=\Omega_0\,J\,.
\end{eqnarray}
Note that in no sense the inhomogeneous equation is the ``dual'' of
the homogeneous one, or vice versa, provided $J\ne 0$.

\subsection{Lorentz force density}

In analogy to Maxwell's theory, we should have in Evans' theory a
Lorentz force density of the type
\begin{equation}\label{Lorentzforce}
f_\a=(e_\a\rfloor \cF^\b)\wedge \cJ_\b\,\quad\mbox{(?)},
\end{equation}
with a Lorentz covariant electric current $\cJ_\a$, which we will
discuss below. However, we didn't find a corresponding definition in
Evans' work.\footnote{In his rebuttal Evans states that ``In truth the
  Lorentz force equation has been obtained ... from the transformation
  properties of the field form $F$...'' Thus Evans claims to derive a
  {\em force} equation of the type (\ref{Lorentzforce}) from the
  transformation properties of the field $\cF^\a$. This is really new
  physics.} Hence we marked this formula by a question mark.

\subsection{``Homogeneous'' field equation of extended electromagnetism}

The exterior covariant derivative of the extended field strength
(\ref{fieldF}) reads
\begin{eqnarray}\label{hom}
  \underbrace{D\cF^\a}_{\rm cov.}=\underbrace{R_\b{}^\a\wedge 
    \cA^\b}_{\rm cov.}
  \qquad\mbox{or}\qquad
  \underbrace{d\cF^\a}_{\rm not\;cov.}=
  \underbrace{R_\b{}^\a\wedge \cA^\b}_{\rm cov.}-
\underbrace{\Gamma_\b{}^\a\wedge \cF^\b}_{\rm not\;cov.}\,;
\end{eqnarray}
here {\it cov.}\ stands for ``covariant under Lorentz rotations of the
{\it frame}'' and {\it not cov.}\ for the lack of that.  This {\it
  Ricci identity for} $\cA^\a$ poses in Evans' unified field theory as
the extension of the homogeneous Maxwell equations.
Eq.(\ref{hom})$_1$ is the analog of the Maxwellian $dF=0$.

If we follow Evans and substitute Evans' ansatz (\ref{ansatz}) into
the right-hand-side of (\ref{hom})$_2$, we have
\begin{eqnarray}\label{hom'}
\underbrace{ d\cF^\a}_{\rm not\;cov.}=
\underbrace{\Omega_0\,\cJ^\a_{\rm hom}}_{\rm not\;cov.}\,,
\end{eqnarray}
with what Evans \cite{Evanspaper50} calls the homogeneous current
\begin{eqnarray}\label{homJ}
  \underbrace{  \cJ^\a_{\rm hom}}_{\rm not\; cov.}:=\frac{a_0}{\Omega_0}
  \left( R_\b{}^\a\wedge \vt^\b-\Gamma_\b{}^\a\wedge T^\b\right) \,.
\end{eqnarray}
Eq.(\ref{hom})$_1$ coincides with Evans \cite{Evans2005a}, Eq.(20).
However, \cite{Evans2003}, Eq.(29), which is also claimed to represent
the homogeneous equation, seems simply wrong.  We are not sure why
Evans substitutes his ansatz only into the right-hand-side of
(\ref{hom})$_2$ and not completely into the whole equation, but this
is just the way he did it in order to find his field equation.

It is strange that the ``current'' (\ref{homJ}) depends on the torsion
and thus on the extended electromagnetic field strength itself:
$T^\b=\cF^\b/a_0$. Thus one cannot specify a current before the
extended electromagnetic field is known. Moreover, this current is not
covariant under Lorentz rotations of the frames since its
right-hand-side depends on the connection explicitly. In contrast, the
conservation law
\begin{equation}\label{pseudocon}
\underbrace{d\cJ^\a_{\rm hom}}_{\rm cov.}=0\,,
\end{equation}
which follows from (\ref{hom'}), is covariant under coordinate
transformations and Lorentz rotations of the frames.  Whereas the
whole equation (\ref{hom'}) is covariant under Lorentz rotations of
the frames, as we recognize from (\ref{hom})$_1$, its left-hand-side
and its right-hand-side for themselves are {\it not} Lorentz
covariant. 

We differentiate (\ref{hom})$_1$ covariantly and recall the second
Bianchi identity (\ref{Bianchi2}):
\begin{equation}\label{xxx}
DD\cF^\a=R_\b{}^\a\wedge \cF^\b\,.
\end{equation}
For reasons unknown to us, Evans \cite{Evans2005a}, p.442, calls this
equation ``the generally covariant wave equation''. If
$R_\b{}^\a\wedge \cF^\b=0$ --- this corresponds to 4 conditions --- he
speaks of the condition for independent fields (no mutual interaction
of gravitation and electromagnetism).

If (i) the curvature vanishes, $R_\b{}^\a=0$, and (ii) the frames are
suitably chosen, $\Gamma_\b{}^\a \stackrel{*}{=}0$, then the field
equation (\ref{hom}) of Evans' theory is really homogeneous: $
d\cF^\a\stackrel{*}{=}0$.  Otherwise we have to live with
inhomogeneous terms. However, Evans claims the following
(\cite{Evans2005a}, p.440): {\it Experimentally} it is found that
(\ref{hom})$_2$ ``must split into the particular solution''
\begin{eqnarray}\label{inh'}
  d\cF^\a&=&0\,,\\ \label{inh''}
  \Gamma_\b{}^\a\wedge \cF^\b&=&R_\b{}^\a\wedge \cA^\b\,.
\end{eqnarray}
Clearly,
Eqs.(\ref{inh'}) and (\ref{inh''}) represent an {\it additional}
assumption.  But note, neither (\ref{inh'}) nor (\ref{inh''}) is
covariant under local Lorentz rotations of the frame.

Since torsion is proportional to the extended electromagnetic field
strength, the first Bianchi identity (\ref{Bianchi1}) and its
contractions (\ref{DC}) and (\ref{DCar}) are alternative versions of
(\ref{hom}), provided one substitutes (\ref{ansatzF}). Eq.(\ref{DCar})
then reads
\begin{equation}\label{homcontr}
  D\left(\eta_{\a\b\g}\wedge\cF^\g \right)+2a_0\,\vt_{[\a}\wedge G_{\b]}=0\,.
\end{equation}

Now the Evans ansatz is exploited and, in order to get the extension
of the inhomogeneous Maxwell equation, Evans had to invest a new idea.

\subsection{Inhomogeneous field equation of extended electromagnetism}

According to our evaluation, Evans' recipe amounts simply to take
the homogeneous equation (\ref{hom})$_1$ and to apply to it the
ad hoc {\em substitution rule}
\begin{equation}\label{subst}
\cF^\a\rightarrow {}^\star\!\cF^\a\quad\mbox{and}\quad
R_\a{}^\b\rightarrow {}^\star \!R_\a{}^\b\,.
\end{equation}
Then one finds
\begin{equation}\label{inhA}
  \underbrace{D\,^\star\!\cF^\a}_{\rm cov.}= \underbrace{{}^\star
    \!R_\b{}^\a\wedge\cA^\b}_{\rm cov.}\qquad\mbox{or}\qquad
  \underbrace{  d\,^\star\!\cF^\a}_{\rm not\;cov.}=
  \underbrace{{}^\star\!R_\b{}^\a\wedge\cA^\b}_{\rm cov.}-\underbrace{
\Gamma_\b{}^\a\wedge \,^\star\!\cF^\b}_{\rm not\;cov.}\,.
\end{equation}
The substitution rule (\ref{subst}) cannot be derived from any
structure in Evans' theory. It represents an additional
asuumption. In particular we stress that $D{}^\star\cF^\a\ne{}^\star
D\cF^\a$, in contrast to Evans' contention. Since $\cF^\a$ as well as
$R_\a{}^\b$ are both 2-forms, the recipe is consistent.
Eq.(\ref{inhA})$_1$ is the analog of the sourceless inhomogeneous
Maxwell equation $d\,^\star\!F=0$.

We substitute the ansatz (\ref{ansatz}) only into the right-hand-side
of (\ref{inhA}) and find
\begin{equation}\label{inh}
 \underbrace{d\,^\star\!\cF^\a}_{\rm not\;cov.}=
 \underbrace{\Omega_0\,\cJ^\a_{\rm inh}}_{\rm not\;cov.}\,,
\end{equation}
with the inhomogeneous current
\begin{equation}\label{inhcurrent}
 \underbrace{  \cJ^\a_{\rm inh}}_{\rm not\;cov.}:=\frac{a_0}{\Omega_0}\left( 
    {}^\star\! R_\b{}^\a\wedge 
    \vt^\b-\Gamma_\b{}^\a\wedge {}^\star T^\b\right)\,.
\end{equation}
Thus, this current is not a vector under Lorentz rotations of the
frame. It has an inhomogeneous tranformation law under Lorentz
rotations of the frame (similar as a connection). If it vanishes in
one frame, it can be non-vanishing in another frame.

Evans \cite{Evanspaper50} claims that (\ref{inh}) can be derived from
(\ref{hom'}) by applying the Hodge star to (\ref{hom'}).  However,
this is not possible. Inter alia, he supposes erroneously that
$^\star\!  d\cF^\a=d\,^\star\!\cF^a$, see also the slides of Eckardt
\cite{workshop}. The inhomogeneous equation represents a new
assumption that can be made plausible by the substitution rule
(\ref{subst}).

As with the homogeneous current, we have again a conservation
law
\begin{equation}\label{pseudoinh}
\underbrace{d\cJ^\a_{\rm inh}}_{\rm cov.}=0\,,
\end{equation}
which is also covariant under Lorentz rotations of the frame.

If we write the inhomogeneous field equation in analogy to the
inhomogeneous Maxwell equation with source $d\,^\star\!F=J$, we have
\begin{equation}\label{inhEvans}
  D\,^\star\!\cF^\a=\cJ^\a\,\qquad\mbox{with}\qquad \cJ^\a:=a_0\,
  {}^\star\! R_\b{}^\a\wedge\vt^\b\,.
\end{equation}
The Lorentz covariant current $\cJ^\a$ seems to be the only current
that could enter the definition (\ref{Lorentzforce}) of the Lorentz
force density. The currents $\cJ^\a_{\rm hom}$ or $\cJ^\a_{\rm inh}$
don't seem to qualify because of their lack of being Lorentz
covariant; see, however, the next section. We differentiate $\cJ^\a$
covariantly:
\begin{equation}\label{nonconserv}
  D\cJ^\a=a_0\left[(D\,^\star\!R_\b{}^\a)\wedge\vt^\b+\,
^\star\! R_\b{}^\a\wedge\,^\star T^\a\right]\,.
\end{equation}
It is not conserved (similarly, as energy-momentum is not conserved in
general relativity). {\it Local electric charge conservation} of
classical electrodynamics $dJ=0$ (note that we have only an exterior
derivative here) is substituted by the four extended charge
non-conservation laws (\ref{nonconserv}).  Local electric charge
conservation, a law that is experimentally established to a high
degree of accuracy (see Particle Data Group \cite{PDG}, p.91, and also
L\"ammerzahl \cite{Lammerzahl}), {\it is irretrievably lost} since the
connection $\Gamma_\a{}^\b$ as well as the torsion $T^\a$ and the
curvature $R_\a{}^\b$ get involved in (\ref{nonconserv}). In Maxwell's
theory no such thing happens for $dJ=0$.

\subsection{Lorentz force density revisited}

We discussed the Lorentz force density earlier, see
(\ref{Lorentzforce}), since it represents the key formula for the
operational definition of the electromagnetic field strength. This
should be also true in Evans' framework. After all, without providing
the action of the electromagnetic field on matter (Lorentz force), the
theory of Evans is simply useless. Evans supplied no
corresponding formula and, accordingly, his field strength $\cF^\a$
has no operational support. However, after defining the
homogeneous and the inhomogeneous currents, the following
observation\footnote{\;I owe this observation to Robert G.\ Flower
  (private communication).  It is also mentioned in Eckardt's workshop
  slides \cite{workshop}, as I found out later.} helps:

The homogeneous current $\cJ_\a^{\rm hom}$ of Evans is of a magnetic
type, whereas $\cJ_\a^{\rm inh}$ is of an electric type. Now we recall
that in Maxwell's theory, if an independent magnetic current 3-form
$K$ is allowed for, the Maxwell equations read
\begin{equation}\label{newmaxwell}
dH=J\,,\qquad dF=K\,,
\end{equation}
compare (\ref{Maxwell1}). If the Lorentz force density is adapted to
this new situation, then we find, see Kaiser \cite{Kaiser} and
\cite{Magnetic},
\begin{equation}\label{inter1}
  f_\alpha = (e_\alpha\rfloor F)\wedge J -  (e_\alpha\rfloor
  H)\wedge K\,.
\end{equation}
Let us translate this into Evans' framework,
\begin{equation}
F\rightarrow\cF^\a\,,\quad K\rightarrow\Omega_0\cJ_\a^{\rm
  hom}\,,\quad H\rightarrow\frac{1}{\Omega_0}\,^\star\cF^\a\,,\quad
J\rightarrow\cJ_\a^{\rm inh}\,,
\end{equation}
that is,
\begin{equation}\label{Lorentzforcenew}
  f_\a=(e_\a\rfloor \cF^\b)\wedge \cJ_\b^{\rm inh}
      -(e_\a\rfloor {}^\star \cF^\b)\wedge  \cJ_\b^{\rm hom}\,.
\end{equation}
We substitute the currents (\ref{inhcurrent}) and (\ref{homJ}):
\begin{eqnarray}\label{Lorentzforcenew'}
  f_\a&= & \frac{a_0}{\Omega_0}\left[(e_\a\rfloor \cF^\b)\wedge\left(
 {}^\star\! R_{\g\b}\wedge 
    \vt^\g-\Gamma_{\g\b}\wedge {}^\star T^\g  \right)\right.\nonumber\\
&&\left.\hspace{13pt}  -(e_\a\rfloor {}^\star \cF^\b)\wedge \left( R_{\g\b}\wedge 
    \vt^\g-\Gamma_{\g\b}\wedge T^\g\right)\right] \,.
\end{eqnarray}
The noncovariant, connection dependent terms on the right-hand-side of
(\ref{Lorentzforcenew'}) drop out, provided we substitute the Evans
ansatz $ \cF^\a=a_0\,T^\a$. We are left with
\begin{equation}\label{Lorentzforcenew''}
  f_\a=\frac{a_0}{\Omega_0}[(e_\a\rfloor
    \cF^\b)\wedge\underbrace{ {}^\star\! R_{\g\b}\wedge 
    \vt^\g}_{\rm el.type\,cur.}
  -(e_\a\rfloor {}^\star \cF^\b)\wedge \underbrace{ R_{\g\b}\wedge 
    \vt^\g}_{\rm mg.type\,cur.}] \,.
\end{equation}
This formula fills the bill. The currents are those on the
right-hand-sides of the covariantly extended Maxwell equations
(\ref{inhA})$_1$ and (\ref{hom})$_1$, respectively.

In our understanding, Eq.(\ref{Lorentzforcenew''}) represents the
Lorentz force formula in Evans' theory. At the same time,
Eq.(\ref{Lorentzforcenew''}) supports our earlier conclusions that
$\cJ_\a^{\rm hom}$ and $\cJ_\a^{\rm inh}$, being non-covariant under
Lorentz rotations of the frames, should not have a fundamental meaning
in Evans' theory. The ``real currents'' can only be read off from the
right-hand-sides of the covariant extended electromagnetic field
equations (\ref{inhA})$_1$ and (\ref{hom})$_1$.

\section{Gravitation: Evans adopted Einstein-Cartan theory of gravity}

\subsection{First field equation of gravity}

According to Evans, the Einstein equation of general relativity needs
to be generalized such the on the left-hand-side we have an asymmetric
Einstein tensor based on RC-geometry and on the right-hand-side an
asymmetric canonical energy-momentum tensor, see \cite{EvansBookI},
p.103, Eq.(5.31). Then his generalized Einstein equation, valid for a
spacetime obeying a RC-geometry, reads (in exterior calculus)
\begin{equation}\label{1st'}
G_\a=\kappa\,\Sigma_\a\,\qquad\qquad\mbox{(first field eq.)},
\end{equation}
where $G_\a$ is the Einstein 3-form (\ref{Ein}) and $\kappa:=8\pi
G/c^3$ (called $k$ by Evans), with $G$ as Newton's gravitational
constant and $c$ the velocity of light.  According to Evans, we have
to understand ${\Sigma}_{\a}$ as canonical energy-momentum that ``has
an antisymmetric component representing canonical angular
energy\footnote{Whatever angular energy may mean in this
  context.}/angular momentum'' (see Evans \cite{Evans2005a}, p.\ 437).
Thus, we take the antisymmetric piece of (\ref{1st'}),
\begin{equation}\label{Riccanti}
\vt_{[\a}\wedge G_{\b]}=\kappa\,\vt_{[\a}\wedge \Sigma_{\b]}\,.
\end{equation}

\subsection{Second field equation of gravity}

It is known from special relativistic field theory, see Corson
\cite{Corson}, Eq.(19.23a), that angular momentum conservation, with
the canonical spin angular momentum current of matter $\tau_{\a\b}$
and the canonical energy-momentum current of matter $\Sigma_\a$, can
be expressed as\footnote{\;Corson \cite{Corson} formulates angular
  momentum conservation in tensor calculus in Cartesian coordinates as
  $\partial_k\frak{S}_{ij}{}^k -2\frak{T}_{[ij]}=0$.  Here
  $i,j,...=0,1,2,3$ are holonomic coordinate indices and
  $\frak{S}_{ij}{}^k$ and $\frak{T}_{ij}$, in Corson's notation,
  canonical spin angular momentum and canonical energy-momentum,
  respectively. If we define the 3-forms of spin and energy-momentum
  as $\tau_{\a\b}=\frak{S}_{\a\b}{}^\g\eta_\g$ and
  $\Sigma_\a=\frak{T}_{\a\b}\,\eta^\b$, respectively, and substitute
  the partial by a covariant exterior derivative, then Corson's
  relation can be translated into (\ref{angmom}). Note that
  $\frak{S}_{\a\b}{}^\g$ and $\frak{T}_{\a\b}$ are ordinary tensors
  here, not, however, tensor densities. We use the Gothic $\frak{T}$
  for energy-momentum in order not to confuse it with the $T$ of the
  torsion.}
\begin{equation}\label{angmom}
D\tau_{\a\b}+\vt_{[\a}\wedge\Sigma_{\b]}=0\,.
\end{equation}
In this form the law is also valid in a RC-spacetime, see \cite{PRs}.

Let us now take a look at the contracted first Bianchi identity
(\ref{DCar}). Then (\ref{DCar}) and (\ref{angmom}),
substituted into (\ref{Riccanti}), yield
\begin{equation}\label{2nd'}
  D\left(C_{\a\b}-\kappa\,\tau_{\a\b} \right)=0\,.
\end{equation}

In this derivation, we invested the asymmetric Einstein equation \`a
la Evans (rather \`a la Sciama-Kibble, see below), the generally
accepted angular momentum law, and the contracted first Bianchi
identity.  Consequently, up to a gradient term, we find
\begin{equation}\label{2nd''}
C_{\a\b}=\kappa\,\tau_{\a\b}\,\qquad\qquad\mbox{(second field eq.)}.
\end{equation}
Now we recall Evans' insistence that spin and torsion are equivalent
(rather proportional to each other, we should say). Provided we drop
the gradient term mentioned, we arrive at (\ref{2nd''}) --- and this,
indeed, expresses the proportionality of spin and torsion. Therefore,
we have shown that (\ref{2nd''}), which is sometimes called Cartan's
field equation of gravity, represents {\it a hidden tacit assumption
  of Evans' theory}.  This proportionality between spin and torsion,
which is {\it not} a geometrical property of torsion, but rather the
result of picking (\ref{1st'}) as one field equation for gravity, is
always advocated by Evans in slogans, but never stated in an explicit
formula, as far as I am aware.  Because of the angular momentum law
(\ref{angmom}), it is clear that the spin $\tau_{\a\b}$ in
(\ref{2nd''}) is the spin of all matter, including that of the
electromagnetic field. Similarly, the energy-momentum $\Sigma_{\a}$ in
(\ref{angmom}) and (\ref{1st'}) represents the energy-momentum of all
matter, including that of the electromagnetic field.

Evans states repeatedly that, within his theory, electromagnetism is
an effect of spin. Let us translate that prose into a quantitative
relation. For this purpose we have to resolve the second field
equation\footnote{We multiply (\ref{2nd''}), with $C_{\a\b}$
  substituted according to (\ref{Car}), from the left with
  $e_\d\rfloor$,
\begin{equation}\label{Car2}\nonumber
  \frac 12(e_\d\rfloor \eta_{\a\b\g})\wedge T^\g 
-\frac 12\,\eta_{\a\b\g}\,e_\d\rfloor\wedge T^\g
  =\kappa\,e_\d\rfloor\tau_{\a\b}\,.
\end{equation}
We have $e_\d\rfloor \eta_{\a\b\g}=\eta_{\a\b\g\d}$, see
\cite{PRs,Birkbook}. Moreover, in order to kill the free indices
$\a,\b,\d$, we multiply with $\eta^{\a\b\d\mu}$ and note
$\eta_{\a\b\g}=\eta_{\a\b\g\nu}\,\vt^\nu$:
\begin{eqnarray}\label{Car3}\nonumber
  -\frac 12\,\eta^{\a\b\d\mu}\eta_{\a\b\d\g}\wedge T^\g
  -\frac
  12\,\eta^{\a\b\d\mu}\eta_{\a\b\g\nu}\,\vt^\nu\wedge e_\d\rfloor T^\g
  =-\kappa\,\eta^{\mu\b\g\d}e_\b\rfloor\tau_{\g\d}\,.
\end{eqnarray}
After some algebra with the products of the $\eta$'s, we find
\begin{equation}\label{Car4}\nonumber
T^\a=-\vt^\a\wedge e_\b\rfloor T^\b+\kappa\,
\eta^{\a\b\g\d}e_\b\rfloor\tau_{\g\d}\,.
\end{equation}
We determine the trace $e_\a\rfloor T^\a$ of the last equation and
re-substitute. This yields the desired result.}  (\ref{2nd''}) with
respect to the torsion $T^\g$:
\begin{equation}\label{inverseC}
T^\a=\kappa\,\eta^{\b\g\d\varepsilon}\left[\d^\a_\b(e_\g\rfloor
\tau_{\d\varepsilon})-
%\mbox{\footnotesize{$\frac{1}{4}$}}
\frac 14\,\vt^\a\wedge (e_\b\rfloor e_\g\rfloor 
\tau_{\d\varepsilon})\right]\,.
\end{equation}
Using Evans' ansatz (\ref{ansatzF}), this transforms into a relation
between the extended electromagnetic field $\cF^\a$ and the spin
$\tau_{\g\d}$:
\begin{equation}\label{inverseC'}
  \cF^\a=a_0\kappa\,\eta^{\b\g\d\varepsilon}\left[\d^\a_\b(e_\g\rfloor
    \tau_{\d\varepsilon})-\frac 14\,\vt^\a\wedge (e_\b\rfloor e_\g
    \rfloor\tau_{\d\varepsilon})\right]\,.
\end{equation}
As soon as we have a source with spin, whatever the source may be,
then, as a consequence of Evans' ansatz (\ref{Ansatz}), an extended
electromagnetic field is created via (\ref{inverseC'}).

We would like to stress that (\ref{2nd''}) and (\ref{1st'}) are {\it
  the field equations of the Einstein-Cartan theory of
  gravity}\footnote{Evans calls (\ref{1st'}) generously the ``Evans
  field equation of gravity'', see \cite{EvansBookI}, p.465.} (1961).
In other words, without stating this explicitly anywhere, Evans just
adopted, knowingly or unknowingly, the two field equations of the
Einstein-Cartan theory. This insight makes a lot of his considerations
more transparent.

In the EC-theory, the gravitational field variables are coframe
$\vt^\a$ and Lorentz connection $\Gamma^{\a\b}=-\Gamma^{\b\a}$. The
{\it first\/} field equation corresponds to the variation of a Hilbert
type Lagrangian with respect to the {\it coframe} and the {\it second}
field equation with respect to the {\it Lorentz connection}.
Consequently, the dynamics of $\vt^\a$ and $\Gamma^{\a\b}$ is
controlled by the two field equations (\ref{1st'}) and (\ref{2nd''}).

\subsection{Trace of the first field equation}

The trace of the first field equation (\ref{1st'}) plays a big role in
Evans' publications. Hence we want to determine it exactly. We
multiply (\ref{1st'}) by $\vt^\a$. Then we get a scalar-valued 4-form
with only one independent component:
\begin{equation}\label{trace1st}
\vt^\a\wedge G_\a=\frac 12\,\vt^\a\wedge\eta_{\a\b\g}\wedge R^{\b\g}=
\kappa\,\vt^\a\wedge\Sigma_\a\,.
\end{equation}
After some light algebra, we find the 4-form (recall
$\Sigma_a=\frak{T}_\a{}^\b\eta_\b$)
\begin{equation}\label{trace1st'}
\eta_{\b\g}\wedge R^{\b\g}=
\kappa\,\vt^\a\wedge\Sigma_\a=\kappa\,\frak{T}\eta\,,
\end{equation}
with the trace of the canonical energy-momentum tensor
$\frak{T}:=\frak{T}_\a{}^\a$. By taking its Hodge dual, remembering
(\ref{scalar'}) and $^\star\eta={}^{\star\star}1=-1$, we can put it
into the scalar form
\begin{equation}\label{trace1st''}
R=-\kappa\,\frak{T}\,.
\end{equation}

This is the generalization of Einstein's trace of his field equation
$\widetilde{R}=-\kappa\,\frak{t}$ to the more general case of
EC-theory. With a {\it tilde\/} we denote the {\it Riemannian\/} part
of a certain geometrical quantity (not to be confused with Evans'
Hodge duality symbol). In general relativity, the source of Einstein's
equation is the symmetric Hilbert energy-momentum tensor
$\frak{t}_{\a\b}=\frak{t}_{\b\a}$; its trace we denote by
$\frak{t}:=\frak{t}_\a{}^\a$.  The corresponding 3-form is
$\sigma_\a=\frak{t}_\a{}^\b\eta_\b$.

In order to make a quantitative comparison with general relativity, we
decompose, within a RC-spacetime, the canonical Noether
energy-momentum $\Sigma_\a=\frak{T}_{\a\b}\,\eta^\b$ into the symmetric
Hilbert energy-momentum $\sigma_\a$ and spin dependent terms according
to \cite{PRs}
\begin{equation}\label{5.6.15} 
\Sigma_{\a} =  
 \sigma_{\a} - e_{\b}\rfloor
(T^{\b}\wedge\mu_{\a}) + D\mu_{\a}\,.
\end{equation}
The spin energy potential $\mu_\a$, a 2-form, is related to the spin
angular momentum 3-form as follows:
$\tau^{\a\b}=\vt^{[\a}\wedge\mu^{\b]}$. Similarly, we decompose the
curvature scalar $R$ into its Riemannian part $\widetilde{R}$ and
torsion dependent terms. The calculations are quite involved. We defer
them to the Appendix. We end up with the final relation 
\begin{equation}\label{tracefinal}
 \widetilde{R}^{\alpha\beta}\wedge\eta_{\alpha\beta}=\kappa\,
\left( \vt^\a\wedge\sigma_\a+ K^{\a\b}\wedge\tau_{\a\b} \right)\,.
\end{equation}
Here $K^{\a\b}$ is the contortion 1-form defined in
(\ref{contortion}). The scalar version of (\ref{tracefinal})
reads\footnote{In his rebuttal, see footnote 1, Evans states the
  following: ``Hehl speaks of `the trace of the first field equation'.
  Nowhere in ECE theory is such a trace mentioned, nowhere needed,
  nowhere is it used.'' However, in the subsequent paragraph, in the
  context of his Eq.(25), he speaks about the role that exactly this
  trace plays in his theory.}
\begin{equation}\label{tracefinal'}
\widetilde{R}=-\kappa\left[\frak{t}+{}^\star(\tau_{\a\b}\wedge
  K^{\a\b}) \right]\,.
\end{equation}
Thus we recognize that in the EC-theory the Riemannian piece
$\widetilde{R}$ of the curvature scalar $R$ obeys a relation like in
general relativity, however, the Einsteinian source $\frak{t}$ has to
be supplemented by a {\it spin-contortion term}.

Our trace formula (\ref{tracefinal'}) of the first field equation,
which is an exact consequence of the EC-theory, should be
distinguished from Evans' corresponding hand-waving expression like,
e.g., \cite{Evanspaper54}, Eq.(17). The $T$ in Evans' formula changes
its meaning within that paper several times; moreover, he uses the
``Einstein Ansatz'' $R=-k\,T$ (in our notation
$\widetilde{R}=-\kappa\,\frak{t}$) even though he is in a
RC-spacetime, where (\ref{tracefinal'}) should have been used instead.

\subsection{Energy-momentum and angular momentum laws}

Within the EC-theory, the energy-momentum law reads, see Sciama,
Kibble, and others
\cite{Sciama1,Kibble,Trautman73,RMP,PRs,Obukhov2006b,Trautman},
\begin{equation}\label{energy-m}
D\Sigma_\alpha
 =(e_\alpha\rfloor T^\beta)\wedge\Sigma_\beta+
(e_\alpha\rfloor R^{\beta\gamma})\wedge\tau_{\beta\gamma}\,.
\end{equation}
Evans assumes incorrectly (as did Cartan in his original papers) that
there has to be a zero on the right-hand-side of (\ref{energy-m}), see
Evans \cite{EvansBookI}, p.\ 464.  This basic mistake, which has
far-reaching consequences, if (\ref{energy-m}) is compared with
(\ref{DEin}), apparently induced Cartan to abandon his gravitational
theory in a RC-spacetime.  We recognize from (\ref{energy-m}) that,
instead of a zero, there rather emerge gravitational Lorentz type
forces of the structure {\it mass $\times$ torsion $+$ spin $\times$
  curvature}. Remember, in electrodynamics we have {\it charge
  $\times$ field strength}.

The angular momentum law, as we saw in (\ref{angmom}), keeps its form
as in flat spacetime, namely
\begin{equation}\label{angular-m}
D\tau_{\a\b}+
\vt_{[\a}\wedge \Sigma_{\b]}=0\,.
\end{equation}

\subsection{Evans' wave equation as a redundant structure}

It is puzzling, besides the structure we discussed up to now, Evans
provides additionally a wave equation for the coframe $\vt^\a$. He
derives it, see \cite{EvansBookI}, p.149, Eq.(8.8), from the
gravitational Lagrangian (in his notation)
\begin{equation}\label{LEv}
L_{\rm Ev}=-\frac{c^2}{k}\left[\frac 12\,(\partial_\mu q_\nu^a)(\partial^\mu
    q_a^\nu) +\frac{R}{2}q^a_\nu q^\nu_a\,\right]\,.
\end{equation}
It is astonishing, Evans presupposes a RC-spacetime; nevertheless, he
takes partial derivatives that are not diffeomorphism invariant. In
order to translate (\ref{LEv}) into a respectable Lagrangian, we (i)
substitute the partial by covariant derivatives
$\partial_\mu\rightarrow D_\mu$ in the sense of minimal coupling in
gauge theories (see Ryder \cite{Ryder}), (ii) interpret $D_\mu
q_\nu^a$ as $2D_{[\mu} q_{\nu]}^a$, and (iii) insert a missing
factor\footnote{Evans equates his expression $q^a_\nu\,q^\nu_a$ always
  consistently to 1, whereas 4 is correct, namely
  $e_\a\rfloor\vt^\a=\d^\a_\a=4$. The trace of the unit matrix in 4
  dimensions is 4.} $\frac{1}{4}$. Then we have (in our notation)
\begin{equation}\label{LEv'}
  L_{\rm Ev'}=-\frac{1}{2\kappa}\left(D\vt^\a\wedge
    {}^\star\!D\vt_\a +{}^\star\!R\right)\,.
\end{equation}
With the definition of torsion and with (\ref{scalar'}), we can
rewrite it as
\begin{equation}\label{LEv''}
  L_{\rm Ev'}=-\frac{1}{2\kappa}\left[T^\a\wedge{}^\star
 T_\a+{}^\star(\vt_\a\wedge\vt_\b)\wedge R^{\a\b}\right]\,.
\end{equation}
Of course, our translation of (\ref{LEv}) into (\ref{LEv'}) is
guesswork. It is definitely clear that $\partial_\mu q_\nu^a\,$, and
thus (\ref{LEv}), is not covariant under general coordinate
transformations (diffeomorphisms). Accordingly, we guessed what Evans
may have had in mind when he wrote down (\ref{LEv}).

In any case, Eq.(\ref{LEv}) represents a purely gravitational
Lagrangian; note the appearance of the gravitational constant in it.
Lagrangians of this type have been widely investigated in the
framework of the Poincar\'e gauge theory of gravity, see
\cite{Obukhov2006a,Erice}.  There, in contrast to EC-theory with its
$R$-Lagrangian, propagating torsion occurs. However, for Evans'
theory, (\ref{LEv''}) is an incorrect Lagrangian.  Only if we dropped
the quadratic torsion term, would we recover the generalized Einstein
equation (\ref{1st'}) that Evans used from the very beginning.
Therefore the Lagrangian $L_{\rm Ev'}$ is false.  But it is more, it
is a redundant structure at the same time.

Our argument is independent of the details of our translation
procedure from (\ref{LEv}) to (\ref{LEv'}). Evans' Lagrangian
(\ref{LEv}) depends on the gravitational constant and the only field
variables present are $\vt^\a$ and $\Gamma^{\a\b}$, i.e., it {\it
  is\/} a gravitational field Lagrangian. However, since Evans
postulates the validity of the generalized Einstein equation
(\ref{1st'}) and of the Cartan equation (\ref{2nd''}), the dynamics of
the variable $\vt^a$ is already taken care of by (\ref{1st'}) and
(\ref{2nd''}).  There is no place for a further wave equation.

In the framework of Evans' theory, the subculture that developed
around Evans' wave equation, is largely inconsistent with Evans'
theory proper, the latter of which will be defined exactly in
Sec.\ref{corner}. Apparently, Evans is misunderstanding his own
theory.
\section{Assessment}

\subsection{Summary of the fundamental structure of Evans' theory}

Since the publications of Evans and associates are not very
transparent to us, we distilled from all their numerous papers and books
the ``spirit'' of Evans' theory.\medskip

\noindent{\bf Geometry:}
Spacetime obeys a RC-geometry that can be described by an orthonormal
coframe $\vt^\a$, a metric $g_{\a\b}=\mbox{diag}(+1,-1,-1,-1)$, and a
Lorentz connection $\Gamma^{\a\b}=-\Gamma^{\b\a}$. In terms of these
quantities, we can define torsion and curvature by, respectively,
\begin{eqnarray}\label{torsion*}
  T^\a&:=&D\vt^\a\,,\\ \label{curv'}
  R_\a{}^\b&:=&d\Gamma_\a{}^\b-\Gamma_\a{}^\g\wedge\Gamma_\g{}^\b\,.
\end{eqnarray}
The Bianchi identities (\ref{Bianchi1},\ref{Bianchi2}) and
their contractions (\ref{DCar},\ref{DEin}) follow therefrom.\medskip

\noindent{\bf Electromagnetism:}
Evans' ansatz relates an extended electromagnetic potential to the coframe,
\begin{equation}\label{Ansatz}
\cA^\a=a_0\,\vt^\a\,.
\end{equation}
The electromagnetic field strength is defined according to 
\begin{equation}\label{DefF}
\cF^\a:=D\cA^\a\,.
\end{equation}

The extended homogeneous and inhomogeneous Maxwell equations read in
Lorentz covariant form
\begin{equation}\label{both}
D\cF^\a=R_\b{}^\a\wedge \cA^\b\qquad\mbox{and}\qquad
D^\star\cF^\a={}^\star\!R_\b{}^\a\wedge\cA^\b\,,
\end{equation}
respectively. Alternatively, with Lorentz non-covariant sources and
with partial substitution of (\ref{Ansatz}) and (\ref{DefF}), they can
be rewritten as
\begin{eqnarray}\label{Evanshom'}
  d\cF^\a=\Omega_0\,\cJ^\a_{\rm hom}\,,\quad&&  \cJ^\a_{\rm hom}
:=\frac{a_0}{\Omega_0}\left( R_\b{}^\a\wedge 
\vt^\b-\Gamma_\b{}^\a\wedge T^\b\right) \,,
\\d\,^\star\!\cF^\a=\Omega_0\,\cJ^\a_{\rm inh}\,,\quad&&  \cJ^\a_{\rm inh}
:=\frac{a_0}{\Omega_0}\left( 
{}^\star\! R_\b{}^\a\wedge 
\vt^\b-\Gamma_\b{}^\a\wedge {}^\star\!T^\b\right)\,.\label{Evansinh'}
\end{eqnarray}
 
\noindent{\bf Gravitation:}
Evans assumes the EC-theory of gravity. Thus, the field equations are
those of Sciama \cite{Sciama1,Sciama2} and Kibble \cite{Kibble}, which
were discovered in 1961:
\begin{eqnarray}\label{Sc1*}
  \frac 12\,\eta_{\a\b\g}\wedge
  R^{\b\g}&\!=\!&\kappa\,\Sigma_\a=\kappa\left(\Sigma_\a^{\rm mat}+\Sigma_\a^{\rm
      elmg} 
  \right)\,,\\ \frac 12\,\eta_{\a\b\g}\wedge T^\g
 &\! =\!&\kappa\,\tau_{\a\b}=\kappa\left(\tau_{\a\b}^{\rm mat}+\tau_{\a\b}^{\rm
      elmg} \right)\,.\label{Sc2*}
\end{eqnarray}
Here $\eta_{\a\b\g}=\,^\star\!\left(\vt_\a\wedge\vt_\b\wedge\vt_\g
\right)$. The total energy-momentum of matter plus electromagnetic
field is denoted by $\Sigma_\a$, the corresponding total spin by
$\tau_{\a\b}$.  

\subsection{Five cornerstones define Evans' unified field theory}\label{corner}

In order to prevent misunderstandings, I'd like to define clearly what
I understand as Evans' unified field theory. Such a statement, which
overlaps with the last subsection, seems necessary since there
are numerous inconsistencies and mistakes in Evans' work, see Bruhn
\cite{Bruhn1,Bruhn2,Bruhn3,Bruhn4,Bruhn5,Bruhn6,Bruhn7,Bruhn8,BruhnBianchi,BruhnAklesh} and Rodrigues et al.\
\cite{Rodrigues1,Rodrigues2}, such that it is necessary to distiguish
between the relevant and the irrelevant parts of Evans' articles.
Let me formulate what I consider to be the five cornerstones of Evans'
theory:
\begin{enumerate}
\item Physics takes place in a Riemann-Cartan spacetime, see
  (\ref{torsion*}) and (\ref{curv'}).

\item The extended electromagnetic potential is proportional to the
  coframe, see (\ref{Ansatz}), and the extended electromagnetic field
  strength to the torsion, see (\ref{DefF}).

\item The extended Maxwell equations are given by (\ref{both}).

\item The Einstein equation gets generalized such that on its
  left-hand-side we have the asymmetric Einstein tensor of a
  Riemann-Cartan spacetime and on its right-hand-side, multiplied with
  the gravitational constant, there acts as source the asymmetic
  canonical energy-momentum tensor of the extended electromagnetic
  field plus that of matter, see (\ref{Sc1*}).

\item Torsion is proportional to spin.
\end{enumerate}
\noindent One may wonder what Evans understood exactly as spin.
However, since he specified the canonical energy-momentum tensor under
cornerstone 4, we concluded that he opts likewise for the
corresponding spin angular momentum tensor under cornerstone 5. This
all the more, since Evans \cite{Evans2005a}, p.\ 437, mentioned the
{\it canonical} spin explicitly.  Starting {}from cornerstone 4, we
were able to show, using only the angular momentum law and a piece of
the first Bianchi identity, that cornerstone 5 implies the second
field equation (\ref{Sc2*}).

There is not more than these five cornerstones. Our conclusions in
this paper and the one accompaying it \cite{AssII} are derived only
{}from these 5 cornerstones by the use of the appropriate mathematics.
\medskip

We disregarded the following two main points:\medskip
%\begin{itemize}
%\item 

A) The antisymmetric part of the metric. Evans has some small talk
about it mixed with partially incorrect formulas, see Bruhn
\cite{Bruhn6}. Because of cornerstone 1, an asymmetric metric is
excluded.  Hence we didn't follow this train of thoughts of Evans any
longer.

B) Evans derived a wave equation for the {\it coframe} in a not too
transparent way, see \cite{EvansBookI}, p.149, Eq.(8.5). All the
results in the context of this wave equation we don't consider to
belong to Evans' theory proper, as defined above. Since the
generalized Einstein equation of cornerstone 4, together with
cornerstone 5, rules already the dynamics of the coframe --- after
all, one can find the generalized Einstein equation by variation of
the curvature scalar with respect to the coframe --- there is no
place for a further equation of motion for the coframe.

In the accompanying paper \cite{AssII} we propose a variational
principle for Evans' theory that reproduces the facts mentioned in
cornerstones 1 to 5. In this context, there emerges an additional
piece $D\,{}^\star T_\a$ on the right-hand-side of the generalized
Einstein equation, which, because of $D\,{}^\star T_\a=D\,{}^\star
D\vt_\a$, is, indeed, in the linearized version a wave operator
applied to the coframe. And this structure is reminiscent of those in
Evans' wave equation. However, our result was achieved by just taking
the five cornerstones for granted and by constructing an appropriate
Lagrangian. We didn't use any additional assumption, whereas Evans
introduces his wave equation as an ad hoc structure without consistent
motivation.
%\end{itemize}

\subsection{Points against Evans' theory}

\subsubsection{Electrodynamics is {\em not} universal and thus cannot induce\\
  a non-Riemannian geometry of spacetime}

In gravity the experimentally well established equality of inertial
and gravitational mass $m_{\rm in}=m_{\rm gr}$ is a fundamental
feature. It is the basis of Einstein's equivalence principle and of
the {\it geometric} interpretation of gravity in the framework of
general relativity. The {\it universality} of this feature is
decisive.  Since, according to our present knowledge, all physical
objects carry energy-momentum, the equivalence principle applies
equally well to all of them.

Is there a similar physical effect known in electromagnetism? No, not
to my knowledge. Rather, the decisive features of electromagnetism are
electric charge and magnetic flux conservation (yielding the Maxwell
equations \cite{Birkbook}). And these conservation laws have nothing
to do with spacetime symmetries, whereas energy-momentum, the source
in Einstein's gravitational theory, is related, via Noether's theorem,
to {\it diffeomorphisms} of spacetime or, in special relativity, to {\it
  translations} in spacetime. In the Maxwell-Dirac theory (Maxwell's
theory with a Dirac electron as source), electric charge conservation
emerges due to the $U(1)$ phase (gauge) invariance of the theory, that
is, due to an {\it internal} symmetry (unrelated to external, i.e.,
spacetime symmetries).  Moreover, charge conservation is universally
valid. However, it has nothing to say about electrically and
magnetically {\it neutral} matter, as, e.g., the neutrinos $\nu_{\rm
  e},\nu_{\rm \mu},\nu_{\rm \tau}$, the photon $\gamma$, the gauge
boson $Z$, the neutral pion $\pi^0$, etc.

Evans provides {\it no new insight} into this question. His only
argument is that any ansatz (like his $\cA^\a=a_0\vt^\a$) must be
permitted and only experiments can decide on its validity. However,
Evans' ansatz $\cA^\a=a_0\vt^\a$ {\it presupposes\/} that
electromagnetism, like the coframe $\vt^\a$, is a universal
phenomenon, which it isn't, since neutral matter is exempt {}from it.
The lack of universality of electromagnetism makes its geometrization
a futile undertaking.

This argument is sufficient for me to exclude Evans' theory right
{}from the beginning.  However, some people, like Evans himself, don't
find it so convincing.  Therefore we collect more evidence.

  \subsubsection{Uncharged particles with spin and charged particles
    without spin cause unsurmountable problems for Evans' theory}

  Take a neutrino, say the electron neutrino $\nu_{\rm e}$. It has no
  electric charge ( $< 10^{-14}$ electron charges), no magnetic moment
  ($<10^{-10}$ Bohr magnetons), and no charge radius squared [$<
  (-2.97\; \mbox{ to }\; 4.14)\times 10^{-32}\; \mbox{cm$^2$}$], see
  \cite{PDG}. Hence the $\nu_{\rm e}$ is electromagnetically neutral
  in every sense of the word. But is carries spin $1/2$.
  Consequently, according to Evans' doctrine, see (\ref{inverseC'}),
  it should create an electromagnetic field, But halt, this cannot be
  true! A neutrino creating an electromagnetic field? Even Evans
  abhors such an idea.  And his remedy? For a neutrino we have to put
  $a_0=0$, is Evans' stunning answer to a corresponding question, see
  Evans'
  blog.\footnote{http://www.atomicprecision.com/blog/2007/02/19/elementary-particles-charge-and-spin-of-ece-theory-2/}
  A {\it unified} field theory of {\it geometric} type that switches
  off a coupling constant for a certain type of matter, doesn't it
  lose all credentials?

  Complementary is the charged pion $\pi^\pm$. It carries electric
  charge but {\it no} spin. Evan
  concludes\footnote{http://www.atomicprecision.com/blog/2007/02/19/elementary-particles-charge-and-spin-of-ece-theory/}
  that it cannot carry an electromagnetic field either!

  Of course, according to Evans' ansatz $\cA^\a=a_0\vt^\a$,
  electromagnetism is assumed to be an universal phenomenon. Since
  this assumption is incorrect, Evans' theory must run into
  difficulties for neutral and for spinless matter willy nilly.

  \subsubsection{There doesn't exist a scalar electric charge,
    electric charge conservation is violated}

  In Maxwell's theory the current $J$ integrated over a
  (3-dimensional) spacelike hypersurface $\Omega_3$ yields a
  4-dimensional scalar charge $\int_{\Omega_3}J$. In Evans' theory no
  such structure is available since any current $\cJ^\a$, because it
  is vector-valued, doesn't qualify as an integrand. Evans didn't
  propose a mechanism for solving that problem. Accordingly, in Evans'
  theory, a global electric charge has not been defined so far in a
  Lorentz covariant way.

  By the same token, as was shown in (\ref{nonconserv}), electric
  charge conservation is violated: $D\cJ^\a\ne 0$. Under such
  circumstances even the concept of a test charge is dubious. Charge
  conservation is a law of nature. Exceptions are not known, see the
  experimental results collected by the Particle Data Group
  \cite{PDG}. Therefore Evans' theory grossly contradicts experiment.

  To take Evans' $\cJ^\a_{\rm hom}$ or $\cJ^\a_{\rm inh}$ as a
  substitute for a decent conserved current is impossible, even when
  $d\cJ^\a_{\rm hom}=0$ and $d\cJ^\a_{\rm inh}=0$. They both,
  $\cJ^\a_{\rm hom}$ and $\cJ^\a_{\rm inh}$, depend explicitly on the
  connection and don't transform as vectors under Lorentz rotations of
  the frames. Their physical interpretation, as given by Evans, since
  {\em not} Lorentz covariant, is null and void. 

  \subsubsection{There doesn't exist a well-defined Maxwellian limit,
    the superposition principle is violated}

  According to our considerations in Sec.\ref{ELEKTRO}, we cannot
  extract from the $SO(1,3)$ electrodynamics proposed by Evans in a
  Lorentz covariant way an $O(3)$ sub-electrodynamics, the latter of
  which Evans claims to be a physical theory. Moreover, we have shown
  that the index $\a$ in $\cA^\a$ cannot be compensated in a Lorentz
  covariant way such as to find the Maxwellian potential $A$ in some
  limit. Thus, we have a potential $\cA^\a$ with 16 independent
  components and we don't know what to do with them, provided we
  insist on covariance under Lorentz rotations of the frames.

  Bruhn \cite{Bruhn4} has even shown explicitly that a plane wave in
  Evans' $O(3)$ electrodynamics, if subject to a Lorentz
  transformation, will not be any longer a plane wave. A proof cannot
  be more telling. In addition, Bruhn \cite{Bruhn8} pointed out in
  detail how Evans suppresses the undesired $\cA^0$ component of his
  potential in order to arrive at his $O(3)$ structure, compare also
  Bruhn and Lakhtakia \cite{BruhnAklesh,Aklesh}.

  Wielandt \cite{Wielandt} demonstrated that the superposition
  principle, valid in Maxwell's theory, breaks down in Evans' $O(3)$
  electrodynamics. In a non-linear theory this is inevitable. However,
  the superposition principle cannot even be recovered for small
  amplitudes and under suitable supplementary conditions. In this
  sense, Maxwell's theory as a limiting case seems to be excluded.

  \subsubsection{Evans' theory is not really unified}

  The energy-momentum and spin angular momentum 3-forms of matter
  $\Sigma_\a^{\rm mat}$ and $\tau_{\a\b}^{\rm mat}$, entering the two
  field equations (\ref{Sc1*}) and (\ref{Sc2*}), have to be determined
  form other physical theories, like {}from Dirac's electron theory.
  Thus Evans' theory is not really unified.\bigskip

  On top of these five main counterarguments --- remember that one
  conclusive counterargument is enough to disprove a theory --- we
  were able to formulate a variational principle for Evans' theory:

  \subsubsection{Evans' theory is trivial and collapses to general
    relativity in all physical cases}

  As Obukhov and the author have shown in an accompanying paper
  \cite{AssII}, Evans theory can be characterized by a {\em dimensionless
  constant}
\begin{equation}\label{dimless}
  \xi:=\frac{a_0^2\kappa}{\Omega_0}\,,
\end{equation}
a fact that was apparently overlooked by Evans. If Evans' ansatz for a
unified field theory is to be taken seriously, then certainly one
would expect $a_0$, and thus $\xi$, to be an universal constant that
cannot be adjusted freely (see, however, Evans' treatment of the
neutrino that was discussed above).

We proposed a variational principle \cite{AssII} with a Lagrange
multiplier term that enforces Evans' ansatz. This approach reproduces
all features of Evans' theory. We find two field equations with 10 +
24 independent components, respectively. The second field equation, it
is (\ref{2nd''}) with the spin of the $\cA^\a$ field on its
right-hand-side, is algebraically linear in torsion and can be solved.
In all physical cases, the torsion vanishes completely and, because of
$\cF^\a=a_0\,T^\a$, Evans' extended electromagnetic field vanishes,
too. Consequently, in all physical cases Evans' theory collapses to
the Einstein vacuum field equation.

Probably Evans will argue that he doesn't like our variational
principle and that our principle ammends the inhomogeneous
electromagnetic field equation (\ref{both})$_2$ and the first
gravitational field equation (\ref{Sc1*}) with terms induced by the
Lagrange multiplier. And that these terms are not contained in his
original theory. This is true. However, we have shown a consistent way
(we believe, it is the only way) to include Evans' ansatz
$\cA^\a=a_0\,\vt^\a$ into the the electromagnetic and gravitational
field equations of Evans' theory. If Evans rejects our variational
principle, he will have a problem. If he substitutes his ansatz into
the extended Maxwell equations (\ref{both}), he will get field
equations for $\vt^\a$ and $\Gamma^{\a\b}$, which are of second order
in $\vt^\a$ (basically wave type equations); if he substitutes his
ansatz also into the gravitational field equations (\ref{Sc1*}) and
(\ref{Sc2*}), which, after an elimination prodecure, are also of
second order in $\vt^\a$, how will he guarantee that these two
different sets of wave type equations are consistent with each other?
Clearly, this cannot be guaranteed. However, our Lagrange multiplier
method does guarantee consistency.

We put this point at the end of our list, since this consequence is
{\it not} inevitable. By abolishing a Hilbert type Lagrangian and
going over to a Lagrangian quadratic in torsion and/or in curvature
(``Poincar\'e gauge theory''), one could ameliorate this situation,
see, e.g., Itin and Kaniel \cite{Itin1,Itin2}, Obukhov
\cite{Obukhov2006a}, and Heinicke et al.\ \cite{aether}. However, we
won't do that because the reasons given above exclude an approach \`a
la Evans. Still, for more than 20 years it is known of how to make
torsion a propagating field, see Sezgin and van Nieuwenhuizen
\cite{Sezgin} and Kuhfuss and Nitsch \cite{KuhNitsch}. In Evans'
theory, one could implement such a mechanism. However, then
cornerstone 5, the proportionality between spin and torsion, had to be
given up, a central point in Evans' approach.

\section{Conclusion}

Around the year 2003, Evans grafted his ill-conceived
$O(3)$-electrodynamics on the viable Einstein-Cartan theory of
gravity, calling it a unified field theory. The hybrid that he
created has numerous genetic defects; some of them are lethal.

\subsubsection*{Acknowledgments} I am most grateful to several people
for greatly helping me to understand the physics and the mathematics
of Evans' work and for guiding me to the relevant literature. In
particular, I would like to mention Gerhard W.~Bruhn (Darmstadt),
Robert G.~Flower (Applied Science Associates), Yuri Obukhov
(Cologne/Moscow), and Erhard Wielandt (Stuttgart). Moreover, Arkadiusz
Jadczyk (Toulouse) helped me with detailed and constructive criticism.
Many thanks to all of them. This work has been supported by the grant
HE 528/21-1 of the DFG (Bonn).

\section{Appendix: Decomposing the trace of the first field
  equation}

We start {}from (\ref{5.6.15}), namely 
\begin{equation}\label{5.6.15'} 
\Sigma_{\a} =  
 \sigma_{\a} - e_{\b}\rfloor
(T^{\b}\wedge\mu_{\a}) + D\mu_{\a}\,,
\end{equation}
and {}from $\tau^{\a\b}=\vt^{[\a}\wedge\mu^{\b]}$. The inverse of the
latter relation reads \cite{PRs}
\begin{equation}\label{5.1.24}
\mu_\alpha=-2e_\beta\rfloor \tau_\alpha{}^\beta + \frac 12\,  
\vartheta_\alpha\wedge(e_\beta\rfloor e_\gamma\rfloor\tau^%  
{\beta\gamma})\,,
\end{equation}
and its contraction is
\begin{equation}\label{5.1.24'}
\vt^\a\wedge\mu_\a=2e_\a\rfloor(\vt^\b\wedge\tau_\b{}^\a)\,.
\end{equation}

Now we recall, see (\ref{trace1st'}), that we only need the
contraction of (\ref{5.6.15'}) with $\vt^\a$:
\begin{equation}\label{trace}
  \vt^\a\wedge\Sigma_\a=\vt^a\wedge\sigma_\a+e_\b\rfloor
\left(\vt^\a\wedge\mu_\a\wedge T^\b \right)
-d\left(\vt^\a\wedge\mu_\a \right)\,.
\end{equation}
This will be substituted in (\ref{trace1st'}). By using
(\ref{5.1.24'}), we find
\begin{eqnarray}\label{tracesubst}
R^{\a\b}\wedge \eta_{\a\b}&=&\nonumber\kappa\left\{\vt^\a\wedge\sigma_\a
+2e_\a\rfloor\left[
T^\a\wedge e^\b\rfloor(\vt^\g\wedge\tau_{\g\b})\right]\right.\\&&\left.
-2d\left[e_\a\rfloor(\vt^\b\wedge\tau_\b{}^\a) \right]
  \right\}\,.
\end{eqnarray}
Obviously, we can now eliminate the spin $\tau_{\a\b}$ by contracting
the second field equation (\ref{2nd''}),
\begin{equation}\label{ddt}
  \kappa\,\vt^\b\wedge\tau_{\b\a}=\eta_{\a\b}\wedge T^\b\,,
\end{equation}
and substituting it in (\ref{tracesubst}). This yields
\begin{eqnarray}\label{tracesubst'}
R^{\a\b}\wedge \eta_{\a\b}&=&\nonumber\kappa\,\vt^\a\wedge\sigma_\a
+2e_\a\rfloor\left[
T^\a\wedge e^\b\rfloor(\eta_{\b\g}\wedge T^\g)\right]\\&&
-2d\left[e_\a\rfloor(\eta^\a{}_\b\wedge T^\b) \right]\,.
\end{eqnarray}
Some algebra shows that the second term on the right-hand-side
vanishes and that $e_\a\rfloor(\eta^\a{}_\b\wedge
T^\b)=\vt^\a\wedge{}^\star T_\a$. Thus,
\begin{eqnarray}\label{new}
R^{\a\b}\wedge \eta_{\a\b}=\kappa\,\vt^\a\wedge\sigma_\a
-2d\left(\vt^\a\wedge{}^\star T_\a \right)\,.
\end{eqnarray}
This is a remarkably simple formula. The first term
$\kappa\,\vt^\a\wedge\sigma_a$ is the Einsteinian trace, the second
one represents a correction by torsion and hence by spin.

We can now study the effect of spin on the Riemannian piece
$\widetilde{R}$ of the curvature scalar $R$. For that purpose, we
start {}from the geometrical decomposition formula
\cite{PRs,chh,Obukhov2006a}
\begin{eqnarray}\label{5.9.18}
R^{\alpha\beta}\wedge\eta_{\alpha\beta}& = &
 \widetilde{R}^{\alpha\beta}\wedge\eta_{\alpha\beta}+ K^{\alpha\mu}\wedge
  K_{\mu}{}^{\beta}\wedge\eta_{\alpha\beta}- K^{\alpha\beta}\wedge
  T^{\gamma}\wedge\eta_{\alpha\beta\gamma}\nonumber\\&&% \hspace{50pt}
- 2d(\vartheta^{\alpha}\wedge
  \,^\star T_{\alpha})\,.
\end{eqnarray}
The second and the third terms on the right-hand-side can be
collected. Then,
\begin{eqnarray}\label{5.9.18'}
  R^{\alpha\beta}\wedge\eta_{\alpha\beta} = 
  \widetilde{R}^{\alpha\beta}\wedge\eta_{\alpha\beta}
  - \frac 12\,K^{\alpha\beta}\wedge
  T^{\gamma}\wedge\eta_{\alpha\beta\gamma}- 2d(\vartheta^{\alpha}\wedge
  \,^\star T_{\alpha})\,.
\end{eqnarray}
The latter equation is substituted into (\ref{new}). The derivatives
drop out and we are left with
\begin{equation}\label{hallo}
  \widetilde{R}^{\alpha\beta}\wedge\eta_{\alpha\beta}=
  \kappa\,\vt^\a\wedge\sigma_\a-\frac 12\,\eta_{\a\b\g}
  \wedge T^\g\wedge K^{\a\b}\,.
\end{equation}
Clearly, the second field equation can be re-substituted and we
arrive\footnote{Often exterior calculus is more effective and
  straightforward than tensor calculus. However, when the connection
  is split in a Riemannian and a post-Riemannian piece, then
  computations in tensor calculus are usually more direct and simpler.
  This is also the case in the derivation of (\ref{tracefinal*}) or
  rather of (\ref{tracefinal'}).} at
\begin{equation}\label{tracefinal*}
 \widetilde{R}^{\alpha\beta}\wedge\eta_{\alpha\beta}=\kappa\,
\left( \vt^\a\wedge\sigma_\a+ K^{\a\b}\wedge\tau_{\a\b} \right)\,.
\end{equation}

\begin{footnotesize}

\end{footnotesize}

\begin{thebibliography}{99}
%\medskip
%\begin{footnotesize}

\bibitem{Bishop} R.L.~Bishop and R.J.~Crittenden, {\it Geometry of
Manifolds,} Academic Press, New York (1964).

\bibitem{Blagojevic} M. Blagojevi\'c: {\sl Gravitation and Gauge
Symmetries,\/} Institute of Physics Publishing, Bristol, UK (2002).

\bibitem{Bruhn3} G.W.~Bruhn, {\it No energy to be extracted {}from the
    vacuum}, Physica Scripta {\bf 74} (2006) 535--536.

\bibitem{Bruhn4} G.W.~Bruhn, {\it No Lorentz property of M W Evans'
    O(3)-symmetry law}, Physics Scripta {\bf 74} (2006) 537--538.

\bibitem{Bruhn1} G.W.~Bruhn, {\it On the non-Lorentz invariance of
    M.W.~Evans' O(3)-symmetry law}, 

  arXiv.org/physics/0607186 (3 pages).

\bibitem{Bruhn2} G.W.~Bruhn, {\it The central error of M.W.~Evans' ECE
    theory - a type mismatch}, 

  arXiv.org/physics/0607190 (6 pages).

\bibitem{Bruhn5} G.W.~Bruhn, {\it Refutation of Myron W.\ Evans
    $B^{(3)}$ field hypothesis,}
   
 { http://www.mathematik.tu-darmstadt.de/$\hspace{4pt}\widetilde{\null}\hspace{4pt}$bruhn/B3-refutation.htm}

  \bibitem{Bruhn6} G.W.~Bruhn, {\it Comments on M.W.~Evans' preprint
      Chapter 2: Duality and the Antisymmetric Metric (p.21 - 30)} 

http://www.mathematik.tu-darmstadt.de/$\hspace{4pt}\widetilde{\null}\hspace{4pt}$bruhn/Comment-Chap2.htm

\bibitem{BruhnBianchi} G.W.~Bruhn, {\it Remarks on Evans' 2nd Bianchi
    Identity},

  http://www.mathematik.tu-darmstadt.de/$\hspace{4pt}\widetilde{\null}\hspace{4pt}$bruhn/EvansBianchi.html

\bibitem{Bruhn7} G.W.~Bruhn, {\it Comments on Evans' Duality},

{ http://www.mathematik.tu-darmstadt.de/~bruhn/EvansDuality.html}

\bibitem{Bruhn8} G.W.~Bruhn, {\it ECE Theory and Cartan Geometry},

http://www.mathematik.tu-darmstadt.de/~bruhn/ECE-CartanGeometry.html

\bibitem{BruhnAklesh} G.W.~Bruhn and A.~Lakhtakia, {\it Commentary
      on Myron W.\ Evans' paper ``The Electromagnetic Sector ...''},
   
{ http://www.mathematik.tu-darmstadt.de/$\hspace{4pt}\widetilde{\null}\hspace{4pt}$bruhn/EvansChap13.html}
   
\bibitem{Cartan1922} \'E. Cartan, {\it Sur une g\'en\'eralisation de
  la notion de corbure de Riemann et les espaces \`a torsion,} C.\ R.\
  Acad.\ Sci.\ (Paris) {\bf 174} (1922) 593--595.

\bibitem{cartan1922} \'E. Cartan, {\it On a generalization of the notion
    of Riemann curvature and spaces with torsion}.  Translation of
  \cite{Cartan1922} {}from the French by G.D. Kerlick. In: {\sl
    Cosmology and Gravitation\/}, P.G. Bergmann, V. De Sabbata, eds.
  Plenum Press, New York (1980) Pp.\ 489--491; see also the remarks of
  A.\ Trautman, ibid.\ pp.\ 493--496.

\bibitem{Cartan} \'E. Cartan: {\it On Manifolds with an Affine
    Connection and the Theory of General Relativity}, English
  translation of the French original, Bibliopolis, Napoli (1986).

\bibitem{CartanGoldberg} E.~Cartan, {\it Riemannian Geometry in an
    Orthogonal Frame,} trans. {}from Russian by V.V.~Goldberg, World
  Scientific, Hackensack, NJ (2001), Sec.87.

\bibitem{Corson} E.M.~Corson, {\it Introduction to Tensors, Spinors,
    and Relativistic Wave-Equations,} Blackie, London (1953).

\bibitem{Debever} R.~Debever (ed.), {Elie Cartan -- Albert Einstein,
  Lettres sur le Parall\'elisme Absolu 1929--1932}, original letters
with translations in English, Palais des Acad\'emies, Bruxelles
(1979), also Princeton University Press, Princeton, NJ.

\bibitem{Rodrigues2} A.L.T.~de Carvalho, W.A.~Rodrigues, Jr, {\it
      The non sequitur mathematics and physics of the `New
      Electrodynamics' of the AIAS Group,} Random Operators and
    Stochastic Equations {\bf 9} (2001) 161--206;
    arXiv.org/physics/0302016.

\bibitem{workshop} H.~Eckardt, {\it Slides {}from the first workshop
      on ECE theory}, 
          
    http://aias.us $\rightarrow$ publications $\rightarrow$ Results of
    first workshop
 
\bibitem{Evanspaper50} M.W.~Evans, {\it Solutions of the ECE field
    equations}, paper 50 of Evans' theory; 

  { http://www.aias.us/documents/uft/a50thpaper.pdf}

\bibitem{Evanspaper54} M.W.~Evans, {\it Wave mechanics and ECE
    theory}, paper 54 of Evans' theory;

  { http://www.aias.us/documents/uft/a54thpaper.pdf}

\bibitem{Evanspaper55} M.W.~Evans, {\it Generally covariant dynamics},
  paper 55 of Evans' theory;
 
 { http://www.aias.us/documents/uft/a55thpaper.pdf}

\bibitem{Evans2003} M.W.~Evans, {\it A generally covariant field
    equation for gravitation and electromagnetism}, Foundations of
  Physics Letters {\bf 16} (2003) 369--377.

\bibitem{Evans2005a} M.W.~Evans, {\it The spinning and curving of
spacetime: The electromagnetic and gravitational fields in the Evans
field theory,} Foundations of Physics Letters {\bf 18} (2005)
431--454.

\bibitem{EvansBookI} M.W.~Evans, {\it Generally Covariant Unified
    Field Theory, the Geometrization of Physics Vol.I,} Arima
  Publishing, Suffolk, UK (2005).

\bibitem{EvansBookII} M.W.~Evans, {\it Generally Covariant Unified
    Field Theory, the Geometrization of Physics Vol.II,} Abramis
  Academic Publishing, publisher@abramis.co.uk (2006).

\bibitem{EvansBookIII} M.W.~Evans, {\it Generally Covariant Unified
    Field Theory - The Geometrization of Physics - Volume III,}
  Amazon.com (2006).

\bibitem{EvansNo63} M.W.~Evans and H.~Eckardt, {\it The resonant
    Coulomb law of Einstein Cartan Evans theory,} paper 63 of Evans'
  theory; 

  http://aias.us/documents/uft/a63rdpaper.pdf

\bibitem{Eyraud} Eyraud, H., {\it La th\'eorie affine asym\'etrique du
    champs \'electromagn\'etique et gravifique et le rayonnement
    atomique}, C.\ R.\ Acad.\ Sci.\ (Paris) {\bf 180} (1925) 1245--1248.

\bibitem{Garcia} A.A.~Garcia, F.W.~Hehl, C.~Heinicke and
  A.~Macias, {\it Exact vacuum solution of a (1+2)-dimensional
    Poincar\'e gauge theory: BTZ solution with torsion,} Phys.\ Rev.
  {\bf D67} (2003) 124016 (7 pages); arXiv:gr-qc/0302097.

\bibitem{Goenner} H.F.M.\ Goenner, {\it On the History of Unified
    Field Theories}, Living Rev.\ Relativity {\bf 7} (2004) (cited on
  01 Dec 2006); http://www.livingreviews.org/lrr-2004-2H.Goenner

\bibitem{Gronwald} F.~Gronwald, {\it Metric-affine gauge theory of
    gravity. I: Fundamental structure and field equations,} Int.\ J.\
  Mod.\ Phys. {\bf D6} (1997) 263--304; arXiv.org/gr-qc/9702034.

\bibitem{Erice} F.~Gronwald and F.W.~Hehl, {\it On the gauge aspects
    of gravity,} in {\sl Proc.\ Int.\ School of Cosm.\ \& Gravit.}
  14\({}^{{\rm th}}\) Course: Quantum Gravity. Held in Erice, Italy.
  Proceedings, P.G.\ Bergmann et al.\ (eds.). World Scientific,
  Singapore (1996) pp.\ 148--198; arXiv.org/gr-qc/9602013.

%\bibitem{Hehl2006} F.W.~Hehl, {\it A remark on an ansatz by M.W.~Evans
%    and the so-called Einstein-Cartan-Evans unified filed theory,}
%  arXiv:physics/0612026v1 (6 pages).
%
\bibitem{Kinematics} F.W. Hehl, {\it On the kinematics of the torsion
    of space-time}, Foundations of Physics {\bf 15} (1985) 451--471.

\bibitem{RMP} F.W. Hehl, P. von der Heyde, G.D. Kerlick, and J.M.
  Nester, {\it General relativity with spin and torsion: Foundations
    and prospects}, Rev.\ Mod.\ Phys.  {\bf 48} (1976) 393--416.

\bibitem{HMcCrea} F.W. Hehl and J.D. McCrea, {\it Bianchi identities
    and the automatic conservation of energy--momentum and angular
    momentum in general-relativistic field theories,} Foundations of
  Physics {\bf 16} (1986) 267--293.

\bibitem{PRs} F.W. Hehl, J.D. McCrea, E.W.  Mielke, and Y. Ne'eman:
  {\it Metric-Affine Gauge Theory of Gravity: Field Equations, Noether
    Identities, World Spinors, and Breaking of Dilation Invariance},
  Phys.\ Rep. {\bf 258} (1995) 1--171.

\bibitem{Birkbook} F.W.~Hehl and Yu.N.~Obukhov, {\it Foundations of
    Classical Electrodynamics: Charge, Flux, and Metric},
  Birkh\"auser, Boston, MA (2003).

\bibitem{Magnetic} F.W.~Hehl and Y.N.~Obukhov, {\it Electric/magnetic
    reciprocity in premetric electrodynamics with and without magnetic
    charge, and the complex electromagnetic field,} Phys.\ Lett.
  {\bf A323} (2004) 169--175; arXiv.org/physics/0401083.
  
\bibitem{Okun} F.W.~Hehl and Yu.N.~Obukhov, {\it Dimensions and units
    in electrodynamics,} General Relativity Gravitation {\bf 37}
  (2005) 733--749; arXiv.org/physics/0407022.

\bibitem{AssII} F.W.~Hehl and Yu.N.~Obukhov, {\it An assessment of
    Evans' unified field theory II}, Foundations of Physics, to be
  published (2007); arXiv.org/physics/0703117.
 
\bibitem{chh} C.~Heinicke, {\it Exact solutions in Einstein's theory
    and beyond}, PH.D.\ thesis, University of Cologne (2005).

\bibitem{aether} C.~Heinicke, P.~Baekler and F.W.~Hehl, {\it
    Einstein-aether theory, violation of Lorentz invariance, and
    metric-affine gravity,} Phys.\ Rev. {\bf D72} (2005) 025012 (18
  pages); arXiv.org/gr-qc/0504005.

\bibitem{Horie} K.~Horie, {\it Geometric interpretation of
    electromagnetism in a gravitational theory with torsion and
    spinorial matter}, Ph.D.\ thesis, University of Mainz (1995);
  arXiv.org/hep-th/9601066.
  
\bibitem{Infeld} L.~Infeld, {\it Zur Feldtheorie von Elektrizit\"at
    und Gravitation}, Phys.\ Zeitschr. {\bf 29} (1928) 145--147.

\bibitem{Itin1} Y.~Itin and S.~Kaniel, {\it On a class of invariant
      coframe operators with application to gravity,} J.\ Math.\
      Phys. {\bf 41} (2000) 6318--6340; arXiv.org/gr-qc/9907023.
 
\bibitem{Itin2} Y.~Itin, {\it Energy-momentum current for coframe
    gravity,} Class.\ Quant.\ Grav. {\bf 19} (2002) 173--189;
  arXiv.org/gr-qc/0111036.

\bibitem{Ark} A.~Jadczyk, {\it Vanishing vierbein in gauge theories of
    gravitation,} arXiv.org/gr-qc/9909060 (17 pages).

\bibitem{Kaiser} G.~Kaiser, {\it Energy-momentum conservation in
    pre-metric electrodynamics with magnetic charges,} J.\ Phys.
  {\bf A37} (2004) 7163--7168; arXiv.org/math-ph/0401028.

 \bibitem{Kibble} T.W.B.~Kibble, {\it Lorentz invariance and the
    gravitational field}, J.\ Math.\ Phys. {\bf 2} (1961) 212--221.

\bibitem{KuhNitsch} R.~Kuhfuss and J.~Nitsch, {\it Propagating
    modes in gauge field theories of gravity,} Gen.\ Rel.\ Grav. {\bf
    18} (1986) 1207--1227.

\bibitem{Aklesh} A.~Lakhtakia, {\it Is Evans' longitudinal ghost
      field $B^{(3)}$ unknowable?} Foundations of Physics Letters {\bf
      8} (1995) 183--186.

\bibitem{Lammerzahl} C.~L\"ammerzahl, A.~Macias and H.~Mueller,
  {\it Lorentz invariance violation and charge (non-)conservation: A general
  theoretical frame for extensions of the Maxwell equations,}
  Phys.\ Rev. {\bf D71} (2005) 025007 (15 pages); arXiv.org/gr-qc/0501048. 

\bibitem{McMapping} J.D.~McCrea, F.W.~Hehl, and E.W.~Mielke, {\it
    Mapping Noether identities into Bianchi identities in general
    relativistic field theories of gravity and in the field theory of
    static lattice defects,} Int.\ J.\ Theor.\ Phys. {\bf 29} (1990)
  1185--1206.

\bibitem{Obukhov2006a} Y.N.~Obukhov, {\it Poincar\'e gauge gravity:
      Selected topics,} Int.\ J.\ Geom.\ Meth.\ Mod.\ Phys. {\bf 3}
    (2006) 95--138; arXiv.org/gr-qc/0601090.

\bibitem{Obukhov2006b} Y.N.~Obukhov and G.F.~Rubilar, {\it
      Invariant conserved currents in gravity theories with local
      Lorentz and diffeomorphism symmetry,} Phys.\ Rev. {\bf D74}
    (2006) 064002 (19 pages); arXiv.org/gr-qc/0608064.
 
\bibitem{PDG} Particle Data Group, {\it Review of Particle Physics},
    J.\ Phys. {\bf G33} (2006) 1--1231.

  \bibitem{Pilch} K.~Pilch, {\it Geometrical meaning of the Poincar\'e
      group gauge theory}, Lett.\ Math.\ Phys. {\bf 4} (1980) 49-51.

  \bibitem{Post} E.J.~Post, {\it Formal Structure of Electromagnetics
      -- General Covariance and Electromagnetics}, North Holland,
    Amsterdam (1962) and Dover, Mineola, New York (1997).

\bibitem{Rodrigues1} W.A.~Rodrigues, Jr. and Q.A.~Gomes de Souza,
    {\it An ambiguous statement called `tetrad postulate' and the
      correct field equations satisfied by the tetrad fields,} Int.\
    J.\ Mod.\ Phys. {\bf D14} (2005) 2095-2150; arXiv.org/math-ph/0411085.
 
\bibitem{Ruggiero} M.L.~Ruggiero and A.~Tartaglia, {\it
      Einstein-Cartan theory as a theory of defects in space-time }
    Am.\ J.\ Phys.  {\bf 71} (2003) 1303--1313.

  \bibitem{Ryder} L.H.~Ryder, {\it Quantum Field Theory}, Cambridge
    University Press, Cambridge, UK (1996).

  \bibitem{Schouten} J.A.~Schouten, {\it Ricci Calculus}, 2nd ed.,
    Springer, Berlin (1954).

\bibitem{Schouten*} J.A.~Schouten, {\it Tensor Analysis for
    Physicists,} 2nd ed.\ reprinted, Dover, Mineola, New York (1989).

\bibitem{Sciama1} D.W.~Sciama, {\it On the analogy between charge and
    spin in general relativity}, in: {\it Recent Developments of
    General Relativity}, Pergamon, London (1962) pp.\ 415--439.

  \bibitem{Sciama2} D.W. Sciama, {\it The physical structure of
      general relativity}, Rev.\ Mod.\ Phys. {\bf 36} (1964) 463--469;
    1103(E).

 \bibitem{Sezgin} E.~Sezgin and P.~van Nieuwenhuizen, {\it New Ghost
      Free Gravity Lagrangians With Propagating Torsion}, Phys.\ Rev.
    {\bf D21} (1980) 3269--3280.

  \bibitem{Sharpe} R.W. Sharpe, {\it Differential Geometry: Cartan's
      Generalization of Klein's Erlangen Program,} Springer, New York
    (1997).

\bibitem{Tonnelat} M.A.~Tonnelat, {\it La th\'eorie du champ
      unifi\'e d'Einstein et quelques-uns de ses d\'eveloppe\-ments,}
    Gauthier-Villars, Paris (1955).

  \bibitem{Trautman73} A.~Trautman, {\it On the structure of the
      Einstein-Cartan equations,} Symp.\ Math.\ (Academic Press,
    London) {\bf 12} (1973) 139--162.

\bibitem{Trautman} A.~Trautman, {\it Einstein-Cartan theory,} in {\sl
    Encyclopedia of Math.\ Physics,} J.-P.\ Francoise et al., eds.,
  Elsevier, Oxford (2006) pp.\ 189--195; arXiv.org/gr-qc/0606062.

\bibitem{RomualdoEgg} R.~Tresguerres and E.W.~Mielke, {\it
    Gravitational Goldstone fields from affine gauge theory,} Phys.\
  Rev. {\bf D62} (2000) 044004 (7 pages).

\bibitem{Wielandt} E.~Wielandt, {\it The superposition principle of
    waves not fulfilled under M.W.~Evans' O(3) hypothesis}, Phys.\
  Scripta {\bf 74} (2006) 539--540; arXiv. org/physics/0607262.

\bibitem{Wise} D.K.~Wise, {\it MacDowell-Mansouri gravity and Cartan
    geometry,} arXiv.org/ gr-qc/0611154.
 \end{thebibliography}
\end{document}